\documentclass[structabstract]{aa}  

\usepackage{amsmath,stmaryrd}
\usepackage{array}
\usepackage{calc}
\usepackage{enumerate}
\usepackage{epsfig,rotating}
\usepackage{lscape}
\usepackage{multirow}
\usepackage{hyperref}

\usepackage{graphicx}
\usepackage{txfonts}
\usepackage{natbib}  

\newcommand{\bd}{\boldsymbol}
\newcommand{\sm}{\scriptscriptstyle}

\def\plotone#1      {\centering \leavevmode \includegraphics[clip=, width=1.00\columnwidth]{#1}}
\def\plottwo#1#2    {\centering \leavevmode \includegraphics[       width=.45\columnwidth]{#1} \hfil \includegraphics[width=.45\columnwidth]{#2}}

\begin{document}
\title{IRIS: A Generic Three-Dimensional Radiative Transfer Code}

   \author{L.~Ibgui        \inst{1},
           I.~Hubeny       \inst{2},
           T.~Lanz         \inst{3}
          \and
           C.~Stehl{\'e}   \inst{1},
          }

   \institute{LERMA, Observatoire de Paris, CNRS et UPMC, 5 place J.~Janssen, 92195 Meudon, France \\  
              \email{laurent.ibgui@obspm.fr, chantal.stehle@obspm.fr}
          \and 
               Steward Observatory, University of Arizona, 933 North Cherry Avenue, Tucson, AZ 85721 \\ 
              \email{hubeny@as.arizona.edu}
          \and
               Laboratoire J.-L. Lagrange, Universit\'{e} de Nice-Sophia Antipolis, CNRS, Observatoire de la C\^{o}te d'Azur, BP 4229, 06304 Nice, France \\ 
              \email{thierry.lanz@oca.eu}
             }

   \date{Received ; accepted }

 
  \abstract
   {For most astronomical objects, radiation is the only probe of their physical
    properties. Therefore, it is important
    to have the most elaborate theoretical tool to interpret observed spectra or images,
    thus providing invaluable information to build
    theoretical models of the physical nature, the structure, and the evolution of the studied objects.}
   {We present IRIS, a new generic three-dimensional (3D) spectral radiative transfer code that generates synthetic spectra, or images. It can
    be used as a diagnostic tool for comparison with astrophysical observations or laboratory astrophysics experiments.}
   {We have developed a 3D short-characteristic solver that works with a 3D nonuniform Cartesian grid. We have implemented a piecewise
    cubic, locally monotonic, interpolation technique that dramatically reduces the numerical diffusion effect. The code takes into account the velocity
    gradient effect resulting in gradual Doppler shifts of photon frequencies and subsequent alterations of spectral line profiles.
    It can also handle periodic boundary conditions. This first version of the code
    assumes Local Thermodynamic Equilibrium (LTE) and no scattering. The opacities and source functions are specified by the user. In the near future,
    the capabilities of IRIS will be extended to allow for non-LTE and scattering modeling.}
   {IRIS has been validated through a number of tests. We provide the results for the most relevant ones, in particular a searchlight beam test,
    a comparison with a 1D plane-parallel model, and a test of the velocity gradient effect.}
   {IRIS is a generic code to address a wide variety of astrophysical issues applied to different objects or
    structures, such as accretion shocks, jets in young stellar objects, stellar atmospheres, exoplanet atmospheres, accretion disks, rotating stellar
    winds, cosmological structures. It can also be applied to model laboratory astrophysics experiments, such as radiative shocks produced with
    high power lasers.}


   \authorrunning{Ibgui et al.}
   \titlerunning{IRIS, a generic 3D radiative transfer code}
   \maketitle

\section{Introduction}  \label{sec:intro}

Essentially all objects in the Universe have a complicated, three-dimensional,
dynamic structure. Understandably, throughout most of the history of astronomy 
the observed objects had to be modeled using significant restrictions and 
approximations. Among them, a set of assumptions about the global geometry 
of an object always played a pivotal role. For instance, stellar atmospheres were 
typically treated assuming a plane-parallel, horizontally-homogeneous geometry, 
which simplified the problem to one spatial dimension. Similarly, stellar winds,
novae, planetary nebulae, and other extended sources, were modeled using 
an assumption of spherical symmetry, which again renders it a one-dimensional 
problem.

In the last two decades, there was much activity devoted to extend the traditional
modeling techniques to treat structures that are truly 3-dimensional (3D). 
In fact, detailed 3D hydrodynamic simulations have now become essentially routine. A non-exhaustive list
of 3D astrophysical (radiation)(magneto)hydrodynamics (RHD, MHD or RMHD) codes includes ZEUS 
\citep{Stone_et_al_1992a,Stone_et_al_1992b}, ENZO \citep{Bryan_and_Norman_1997,Norman_and_Bryan_1999,OShea_et_al_2004,Norman_et_al_2007},
RAMSES \citep{Teyssier_2002}, HERACLES \citep{Gonzalez_et_al_2007}, PLUTO \citep{Mignone_et_al_2007}, ATHENA \citep{Stone_et_al_2008}
and its radiation module \citep{Davis_et_al_2012}, AstroBEAR \citep{Cunningham_et_al_2009}.

However, since most of the information about an astronomical object is conveyed 
to a distant observer by its radiation, detailed 3D hydrodynamic models
have to be accompanied by adequate 3D radiation transfer solutions
that provide a spectroscopic diagnostic information. Compared to 3D hydrodynamic
models, the 3D radiation transfer solvers are much more complicated and difficult
to treat numerically, because
(i) there are many more quantities that describe
a radiation field, as compared to the hydrodynamics, due to the directional and 
spectral dependence of radiation;
(ii) a long-range interaction between the radiation field and the plasma arises because
of a typically much larger mean free path of a photon compared to a mean free
path of massive particles.

Although providing exact time-dependent, non-LTE radiation (magneto)hydrodynamic models
of astronomical objects is generally viewed as a mighty goal, such a goal has not been fully
achieved yet. However, a large progress was accomplished in recent years, specifically by
3D RMHD codes for stellar atmospheres, such as: the Copenhagen-Oslo Stagger Codes
\citep{Nordlund_and_Galsgaard_1995,Galsgaard_and_Nordlund_1996,Hansteen_2004,Hansteen_et_al_2007},
$\rm CO^5BOLD$ \citep{Freytag_et_al_2002}, MURaM \citep{Voegler_et_al_2005}, 
Bifrost \citep{Gudiksen_et_al_2011}.
All of these codes solve the 3D radiative transfer by approximating the monochromatic opacities 
with a small number of mean opacities (the opacity binning scheme of \citealt{Nordlund_1982}). With
typically four bins, they can model the frequency-integrated radiative energy losses and gains
quite well.

There are also several recently developed computer codes specifically designed
to provide 3D radiative transfer solvers, typically for a post-processing of
3D hydrodynamic simulations.
The most widely used among them are the following:
MUGA \citep{Auer_and_Paletou_1994,Auer_et_al_1994,Trujillo_Bueno_and_Fabiani_Bendicho_1995,Fabiani_Bendicho_et_al_1997},
MULTI3D \citep{Botnen_1997,Leenaarts_and_Carlsson_2009,Leenaarts_et_al_2009}, which is a generalization of the 1D code MULTI
\citep{Carlsson_1986}, the RH code \citep{Uitenbroek_2001,Uitenbroek_2006},
a 3D version of the atmospheric code PHOENIX \citep{Hauschildt_and_Baron_2006,Baron_and_Hauschildt_2007,Hauschildt_and_Baron_2008},
ASSeT \citep{Koesterke_et_al_2008,Koesterke_2009},
SCATE \citep{Hayek_et_al_2011}.
They differ in their intended range of applications, and in many details of the
numerical techniques. They all use some variant of the short-characteristics scheme
(see Section \ref{sec:3D_SC}), but differ in details of what is assumed about the behavior 
of state parameters between the grid points, and how the necessary interpolations
are being performed. For details, the reader is referred to the above mentioned
papers, and several excellent review papers \citep{Carlsson_2008,Carlsson_2009}.

In this paper, we describe our variant of the three-dimensional radiative transfer solver,
named IRIS. Analogously to most of the above mentioned techniques, it uses the 
short-characteristics scheme. Our solver differs from the previous ones in several
respects: it is formulated only in the Cartesian coordinate system, but within this
system it is formulated in the most universal way that allows for several variants
of the spatial integration and interpolation to be tested and used depending on the
actual application; it carefully treats subgriding to allow for line radiation transport
in the presence of arbitrary (non-relativistic) velocity fields. Although the current
version assumes Local Thermodynamic
Equilibrium (LTE) and no scattering, the code is prepared to be relatively
straightforwardly extended to treat scattering and departures from LTE, the so-called
non-LTE (or NLTE) situations. This will be reported in future papers.

The paper is organized as follows. In Section~\ref{sec:RTE} we outline the main physical assumptions of the model solved by IRIS.
In Section~\ref{sec:3D_SC} we summarize the short-characteristics method, and its application in a 3D Cartesian grid.
In Section~\ref{sec:pcw_cubic_mon_intp} we describe the piecewise, cubic, locally monotonic, interpolation technique that we use.
Then, in Section~\ref{sec:propagation_ray} we describe our implementation in IRIS of the techniques presented in Sections~\ref{sec:3D_SC} and
\ref{sec:pcw_cubic_mon_intp}. We also explain the procedure to handle general macroscopic velocity fields.
The calculation of the moments of the radiation field and the related angular integration methods are the subjects of
Section~\ref{sec:moments_angular_integration}.
We detail in Section~\ref{sec:PBC} the method that we employ to treat media with periodic boundary conditions. 
Section~\ref{sec:tests} presents the results of relevant tests.
We conclude our paper in Section~\ref{sec:conclusion} with a summary of the features of IRIS, 
and with an outline of possible extensions to the code.

\section{Radiative Transfer: Basic Definitions and Assumptions}  \label{sec:RTE}

The general unpolarized radiative transfer equation (RTE) in the observer's frame reads
\citep{Chandrasekhar_1950,Mihalas_1978,Mihalas_and_Mihalas_1984}
\begin{multline}  \label{eq:RTE1}
   \left(\frac{1}{c}\ \frac{\partial}{\partial t} + \bd{n}\bd{\cdot}\nabla\right) I(\bd{r},\bd{n},\nu,t) \\
  = \eta(\bd{r},\bd{n},\nu,t) \ - \ \chi(\bd{r},\bd{n},\nu,t)\ I(\bd{r},\bd{n},\nu,t) \:,
\end{multline}

\noindent where $I(\bd{r},\bd{n},\nu,t)$ is the specific intensity of radiation at position $\bd{r}$, propagating in the direction
specified by the unit vector $\bd{n}$, at frequency $\nu$, and time $t$; $\chi(\bd{r},\bd{n},\nu,t)$ is the absorption coefficient,
$\eta(\bd{r},\bd{n},\nu,t)$ is the emission coefficient, and $c$ is the speed of light. Although the above quantities depend on time, the
current version of IRIS solves the time-independent RTE, for a given hydrodynamics structure at a given instant, i.e., for a given snapshot.
In other words,
we consider regimes in which the photon free-flight time is small compared to the fluid flow dynamical timescales, so that the radiation field gets
fully stabilized before any change occurs in the flow dynamical properties. We assume that the typical flow velocities of the hydrodynamic structures
are non relativistic.

As is customary, we introduce the source function,
 \begin{equation} \label{eq:S}
     S(\bd{r},\bd{n},\nu,t) = \frac{\eta(\bd{r},\bd{n},\nu,t)}{\chi(\bd{r},\bd{n},\nu,t)}
 \end{equation}
To simplify the notations, we drop the explicit dependence of radiative quantities
on the position, direction and time, and denote the dependence on frequency 
with a subscript $\nu$.
The transfer equation~(\ref{eq:RTE1}) becomes
   \begin{equation}  \label{eq:RTE2}
       \frac{\partial I_{\nu}}{\partial s} = \chi_{\nu} ( S_{\nu}-I_{\nu}) \:,
   \end{equation}
\noindent where $s$ is the path length along the ray in the direction of propagation of the radiation.

The choice of a numerical method depends on the purpose of  the
simulation. If one is interested only in computing emergent specific
intensity from a 3D computational box, and if the opacity and the 
source function are fully known within the box (the so-called ``formal
solution''), one can select a number of rays (photon paths)
that emerge from the box and cross the whole extent of the box,
and then solve the radiative transfer equation along such rays.
Obviously, a ray does not generally go through the original grid points.
One can choose any discretization of the ray, and then obtain needed
values of the opacity and source function at the discretized positions by
a three-dimensional interpolation. A more efficient way is to discretize
the ray by taking its intersections with the individual planes defined
by the grid points; in this case one deals with two-dimensional 
interpolations. 

The problem is thus reduced to a set of 1D problems. Any method
capable of solving the transfer equation along a ray can be used
here; most popular in the astrophysical applications being  the
Feautrier method, its 2nd order variant \citep{Feautrier_1964} or the 
4-th order variant \citep{Auer_1976}; the Discontinuous Finite Element 
method \citep{Castor_et_al_1992};  or the ``1D short-characteristics'' 
method \citep{Olson_and_Kunasz_1987}. All these methods are
called the ``long-characteristics'' scheme in some specific situations. However, the
terminology is not used consistently; some studies use this term in
a more restricted meaning, namely for using a 1D short-characteristics 
scheme for solving the transport for a set of rays passing through the
whole computational region. 

The situation is different when the source function is not known a priori
within the computational box. The simplest situation of this sort arises 
when the opacity is specified in the computational domain 
and one can even assume LTE, 
but the continuum scattering is taken into account. Another case is 
a general non-LTE situation where line scattering also is important.  
The most complex situation is when dealing with the true radiation 
hydrodynamics, where one has to pass the values of radiation
moments at all grid points to the hydrodynamics part of the code.

Even in the simplest case of LTE with electron (Thomson) scattering,
the total source function depends on the mean intensity of radiation, and
therefore has to be determined iteratively. The essential point is, however,
that even if we are interested only in the emergent radiation, we have to
determine the total source function, and therefore the specific intensities, 
in all grid  points to be able to interpolate it into the discretized positions 
along the rays.

Although the long-characteristics scheme is useful for some
purposes, such as allowing for an efficient parallel scheme
within domain decomposition \citep{Heinemann_et_al_2006}, we chose
the short-characteristics scheme. The classical setup of
the long-characteristics method is to pass a ray through
each grid point through the whole computational box,
in which case the number of necessary operations is obtained
as follows.
Let us assume that the 
computational box is discretized using $N$ points in each direction, 
so that there are $N^3$ grid points (and also $\approx N^3$ elementary 
cells). A long-characteristics scheme would consider  $N^3$ rays. Solving 
the transfer equation along one ray, for one direction and one frequency, 
would require $O(N)$ operations, an exact value depending on how 
exactly are the necessary interpolations performed, assuming that they
require $O(1)$ operations. For one direction and one frequency one thus 
needs $O(N^4)$ operations. In contrast, the short-characteristics scheme 
solves the transfer equation within one cell only, and therefore the total 
number of operations is proportional to the number of cells, that is 
$O(N^3)$ operations.

To avoid confusion, we mention that when using the long-characteristics 
scheme for obtaining an emergent intensity only, it requires also $O(N^3)$ 
operations, because we have $N^2$ rays; each going through one grid
point on the boundary plane that faces an ``observer'', and a transfer 
solution along each ray requires $O(N)$ operations. In this particular case, 
the scaling of the computer time is the same as for the short-characteristics 
scheme. In some cases, the use of long-characteristics scheme may even 
be more computationally advantageous than the short-characteristics, and
therefore some codes (e.g., ASSeT, SCATE) use the short-characteristics 
scheme when dealing iteratively with scattering, and use the 
long-characteristics scheme to evaluate the emergent intensity
for the final converged model.

We note that there are several possible numerical methods that propagate
an information form one edge of the cell to another, whose numbers of operations also scale as
the number of cells, $O(N^3)$. In astrophysics, the vast majority of approaches
is based on the short-characteristics scheme, and this is what we adopt in this 
paper as well. In this paper we assume LTE and no scattering.
In this case, the absorption coefficient $\chi_{\nu}$ is a function of
the local mass density $\rho$ and the local temperature $T$, and the source function
$S_{\nu}$ is equal to the local Planck function $B_{\nu}(T)$.
With these two assumptions (LTE and no scattering), we could have adopted the long-characteristics
method as well, but in view 
of our intended future development of the solver to treat scattering and non-LTE
situations, the short-characteristic scheme is clearly the appropriate choice.

\section{3D Short-Characteristics}  \label{sec:3D_SC}

In the context of astrophysical radiative transfer, the method was first used by
\citet{Mihalas_et_al_1978}, and later by \citep{Olson_and_Kunasz_1987,
Kunasz_and_Auer_1988,Kunasz_and_Olson_1988}, and subsequently
in many studies -- see references below.

\subsection{Overview of the Method}  \label{subsec:overview_SC}
Let us consider, at time $t$, the radiation propagating in direction $\bd{n}$.
The optical depth from position $\bd{r}$ to position $\bd{r}+s\bd{n}$, where $s$ is the path length between these two positions, is:
\begin{equation}  \label{eq:optical_depth}
     \tau(\bd{r},\bd{r}+s\bd{n},\nu,t) = \int_{0}^{s}\chi(\bd{r}+s'\bd{n},\bd{n},\nu,t)\ ds'
\end{equation}

For a radiation propagating from an upwind position $\bd{r_{u}}$ to a current position $\bd{r_{c}}$, the integral form of the formal solution
of the time-independent RTE is:
\begin{equation}  \label{eq:RTE3}
   I_{c} = I_{u} e^{-\tau_{uc}}  +  \int_{0}^{\tau_{uc}} S(\tau) e^{-\left(\tau_{uc}-\tau\right)}\ d\tau \:,
\end{equation}
where $I_{c}$=$I\left(\bd{r_{c}},\bd{n},\nu,t\right)$, $I_{u}$=$I\left(\bd{r_{u}},\bd{n},\nu,t\right)$, $\bd{n}$  is the unit vector along the straight
line $(\bd{r_{u}},\bd{r_{c}})$, $\tau$ is the optical depth from $\bd{r_{u}}$ to an intermediate position $\bd{r_{u}}+s\bd{n}$,
with $0 \leqslant s \leqslant \|\bd{r_{c}}-\bd{r_{u}}\|$, and $\tau_{uc}$ is the optical depth from $\bd{r_{u}}$ to $\bd{r_{c}}$.

\begin{figure} 
\plotone{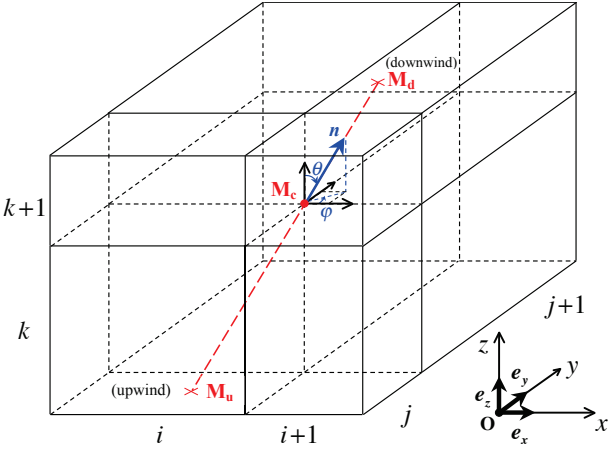}
\caption{Short-characteristics method illustrated with an example of a 3D irregularly spaced Cartesian grid.
         We consider all quantities in the observer's frame. They are defined in the vertices of the cells (grid points).
         $(O,x,y,z)$ is an example of a coordinate system
         with the unit vectors $(\bd{e_{x}},\bd{e_{y}},\bd{e_{z}})$.
         The specific intensity is calculated in the current point~$M_{c}$, for a radiation propagating from the upwind end~point $M_{u}$
         to the downwind end~point $M_{d}$. The direction $\boldsymbol{n}$ of the ray is
         defined by the polar angle $\theta$ and the azimuthal angle $\varphi$. The short-characteristic is
         defined by the line joining $M_{u}$ and $M_{c}$. The transfer equation is solved
         in its integral form along this short-characteristic. The cells around point~$M_{c}$ are
         marked with the following indices: \textit{i}, \textit{i}+1 in \textit{x}-direction, \textit{j}, \textit{j}+1 in
         \textit{y}-direction, and \textit{k}, \textit{k}+1 in \textit{z}-direction.}
\label{fig:SC_global}
\end{figure}

We consider a three-dimensional Cartesian grid (see Figure~\ref{fig:SC_global}) defined in the $(O,x,y,z)$ coordinate system with the unit vectors
$(\bd{e_{x}},\bd{e_{y}},\bd{e_{z}})$. The grid cells are rectangular boxes with irregular spacing in each direction. The sizes of the cells in each
direction depend on the position in this direction: $\Delta x_{\rm{cell}}(x)$, $\Delta y_{\rm{cell}}(y)$, and $\Delta z_{\rm{cell}}(z)$. The
state parameters, i.e., the mass density, the temperature, and the velocity components in the observer's frame, are defined at each grid
point. They are provided by the (radiation)(magneto)hydrodynamics (RMHD) simulations. A direction of propagation of the radiation, defined by the unit
vector $\bd{n}$, is specified by the polar angle~$\theta$ between $\bd{n}$ and the unit vector~$\bd{e_z}$, and by the azimuthal angle~$\varphi$,
between $\bd{e_{x}}$ and the projection of $\bd{n}$ on the $x-y$ plane. In order to span the whole $4\pi$~sr angular domain, we impose
$\theta \in [0,\pi]$ and $\varphi \in [0,2\pi[$.

The task is to calculate the specific intensity $I_{c}$ in a current point~$M_{c}$ in direction $\bd{n}$. 
$M_{c}$ is defined by the intersection of the cells marked with indices \textit{i}, \textit{i}+1 in \textit{x}-direction, \textit{j}, \textit{j}+1 in
\textit{y}-direction, and \textit{k}, \textit{k}+1 in \textit{z}-direction (see Fig.~\ref{fig:SC_global}).
This is accomplished by using Equation~(\ref{eq:RTE3}).
Following the propagation of the ray that goes through $M_{c}$ in direction $\bd{n}$, we define a short-characteristic by the line joining the
intersection of the ray with the upwind cell, $M_{u}$, to the current point $M_{c}$ (the upwind cell is defined by the indices $(i,j,k)$ in
Fig.~\ref{fig:SC_global}).
$I_{c}$ can be determined  if we know the following quantities: the upwind specific intensity $I_{u}$, the
source function $S(\bd{r},\bd{n},\nu,t)$ and the absorption coefficient $\chi(\bd{r},\bd{n},\nu,t)$ (in order to deduct the optical depths $\tau$ and
$\tau_{uc}$) along the short-characteristic from $M_{u}$ to $M_{c}$. However, $S$ and $\chi$ are specified only in the vertices of the cells (grid
points). Therefore, we are essentially free to define laws of variation of these quantities along the
short-characteristic, typically as low-order polynomials; we choose third degree polynomials. To do so, we need to know $(S,\chi)$ in
the upwind end~point~$M_{u}$, $(S_{u},\chi_{u})$, and in the downwind end~point~$M_{d}$, $(S_{d},\chi_{d})$. $M_{d}$ is the intersection of the
short-characteristic with the downwind cell, which is defined by the indices $(i+1,j+1,k+1)$ in Fig.~\ref{fig:SC_global}.
We also choose third degree polynomials to get the upwind quantities $(I_{u},S_{u},\chi_{u})$ and the downwind quantities $(S_{d},\chi_{d})$,
through interpolations from the values in the neighboring grid points. Further, in Section~\ref{sec:pcw_cubic_mon_intp}, we describe and discuss in detail
the discretization of Equation~(\ref{eq:RTE3}), and the mathematical form of the third degree polynomials.

\subsection{Intersections of a Short-Characteristic with the Neighboring Cell Faces}  \label{subsec:intersec_SC}
The coordinates of the intersections of a short-characteristic with the neighboring cell faces, i.e., the coordinates of the upwind end~point $M_{u}$
and the downwind end~point $M_{d}$, are easily determined in the local coordinate system defined by $(M_{c},\bd{e_{x}},\bd{e_{y}},\bd{e_{z}})$
(see~Fig.\ref{fig:SC_global}).

\begin{figure} [ht]
{\centering \leavevmode \includegraphics[width=1.00\columnwidth]{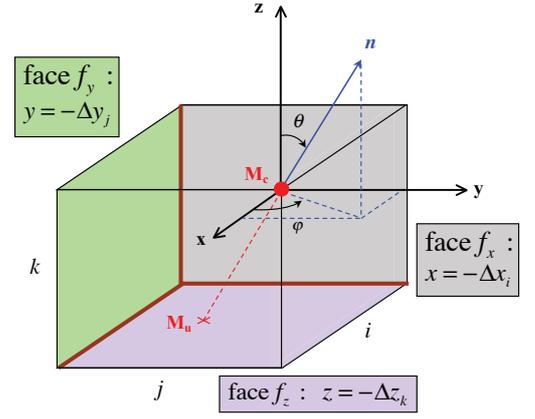}}
{\centering \leavevmode \includegraphics[width=1.00\columnwidth]{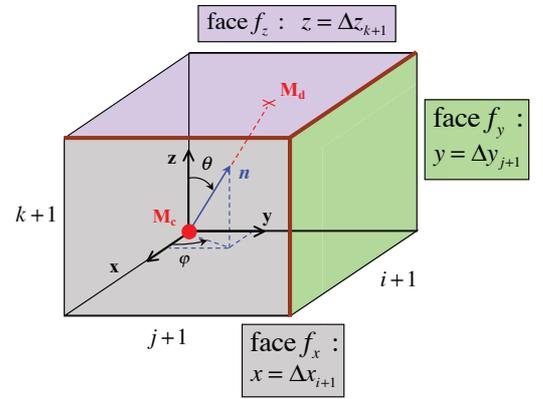}}
\caption{Upwind cell (top) and downwind cell (bottom), for the short-characteristic $M_{u}M_{c}$ displayed in Fig.~\ref{fig:SC_global}.
In this example, the direction of propagation of the radiation, $\bd{n}$, is in the first octant, which corresponds to
$(\theta,\varphi) \in [0,\pi/2[ \times [0,\pi/2[$. The local coordinate system is defined by $(M_{c},\bd{e_{x}},\bd{e_{y}},\bd{e_{z}})$.
The upwind end~point $M_{u}$ belongs to the cell face $f_{z}: z=-\Delta z_{k}$. However, depending on $(\theta,\varphi)$ in this octant,
$M_{u}$ may belong to $f_{x}: x=-\Delta x_{i}$, or to $f_{y}: y=-\Delta y_{j}$.
In the general case, $\Delta x_{i} \ne \Delta y_{j}$, $\Delta x_{i} \ne \Delta z_{k}$, $\Delta y_{j} \ne \Delta z_{k}$, though equalities are possible.
The downwind end~point $M_{d}$ belongs to $f_{z}: z=\Delta z_{k+1}$. However, depending on $(\theta,\varphi)$
in this octant, $M_{d}$ may belong to $f_{x}: x=\Delta x_{i+1}$ or to $f_{y}: y=\Delta y_{j+1}$. See
Section~\ref{subsec:intersec_SC} for more details.}
\label{fig:SC_local}
\end{figure}

Figure~\ref{fig:SC_local} displays the upwind cell (top panel) and the downwind cell (bottom panel), as they are defined for the short-characteristic
plotted in Fig.~\ref{fig:SC_global}. In this example, the upwind cell is determined by indices $(i,j,k)$, while the downwind cell is determined by
indices $(i+1,j+1,k+1)$.
Let us refer to $f_{x}$ (respectively $f_{y}$, $f_{z}$) as the generic name of a cell face that is perpendicular to $x$-axis (respectively $y$-axis,
$z$-axis), and that may be intersected by a short-characteristic, i.e., to which $M_{u}$ or $M_{d}$ may belong.
In our example, $f_{x}$ in the upwind cell is specified by $x=-\Delta x_{i}$, while $f_{x}$ in the downwind cell is specified by $x=\Delta x_{i+1}$,
both in the local coordinate system. In the same vein, $f_{y}$ is $y=-\Delta y_{j}$ in the upwind cell, and $y=\Delta y_{j+1}$ in the downwind cell.
Finally, $f_{z}$ is $z=-\Delta z_{k}$ in the upwind cell, and $z=\Delta z_{k+1}$ in the downwind cell.

In the example displayed in Fig.~\ref{fig:SC_global} and~\ref{fig:SC_local}, $M_{u}$ belongs to the face $f_{z}: z=-\Delta z_{k}$, and $M_{d}$ belongs
to the face $f_{z}: z=\Delta z_{k+1}$. However, depending on the values of $(\theta,\varphi)$, $M_{u}$ may belong to $f_{y}: y=-\Delta y_{j}$, or
to $f_{x}: x=-\Delta x_{i}$, and $M_{d}$ may belong to $f_{y}: y=\Delta y_{j+1}$, or to $f_{x}: x=\Delta x_{i+1}$. In addition, the upwind cell and
the downwind cell are respectively $(i,j,k)$ and $(i+1,j+1,k+1)$ only for $(\theta,\varphi) \in [0,\pi/2[ \times [0,\pi/2[$, as exemplified in
Fig.~\ref{fig:SC_global} and \ref{fig:SC_local}. In fact, for each grid point,
eight different upwind cells are possible, one different cell for $(\theta,\varphi)$ in each of the eight octants in the $4\pi$~sr directional domain.
Let us refer to $\Delta x_{u}$ and $\Delta x_{d}$ as the generic names for the size in $x$-direction of, respectively, the upwind and the downwind
cell, for a given current point $M_{c}$ and a given direction $(\theta,\varphi)$. Similarly, let us introduce the analogous notations
$\Delta y_{u}, \Delta y_{d}$ and $\Delta z_{u}, \Delta z_{d}$ for the corresponding sizes in $y$ and $z$ directions.
Table~\ref{tab:octant_u} indicates, for each of the eight octants, the corresponding range of $(\theta,\varphi)$, the indices of the
upwind cell, along with the values of $\Delta x_{u},\Delta y_{u},\Delta z_{u}$, for any current grid point $M_{c}$.
Table~\ref{tab:octant_d} indicates the equivalent quantities for the downwind cell.

The following generic inequalities identify the cell face to which an upwind (respectively downwind) end~point belongs:
\begin{subequations}  \label{eq:u_d_endpoints}
\begin{alignat}{2}
   M_{u(d)} \in f_{x} &\Leftrightarrow
   |\tan \varphi | < \frac{\Delta y_{u(d)}}{\Delta x_{u(d)}},  |\tan \theta \cos \varphi | \ge \frac{\Delta x_{u(d)}}{\Delta z_{u(d)}} \\
   M_{u(d)} \in f_{y} &\Leftrightarrow
   |\tan \varphi | \ge \frac{\Delta y_{u(d)}}{\Delta x_{u(d)}},|\tan \theta \sin \varphi | \ge \frac{\Delta y_{u(d)}}{\Delta z_{u(d)}} \\
   M_{u(d)} \in f_{z} &\Leftrightarrow
   |\tan \theta \cos \varphi | < \frac{\Delta x_{u(d)}}{\Delta z_{u(d)}}, |\tan \theta \sin \varphi | < \frac{\Delta y_{u(d)}}{\Delta z_{u(d)}}
\end{alignat}
\end{subequations}

The local coordinates of the upwind (respectively downwind) end~point $M_{u(d)}$ depend on the face it belongs to, as follows:
\begin{subequations}  \label{eq:u_coord}
\begin{alignat}{2}
   \underset{\in f_{x}}{M_{u(d)}}:
   \begin{cases}
   x_{u(d)}=\alpha_{xu(d)} \Delta x_{u(d)}                                 \\
   y_{u(d)}=\alpha_{xu(d)} \Delta x_{u(d)} \tan \varphi                    \\
   z_{u(d)}=\alpha_{xu(d)} \frac{\Delta x_{u(d)}}{\tan \theta \cos \varphi}\\
   \end{cases}
\end{alignat}
\begin{alignat}{2}
   \underset{\in f_{y}}{M_{u(d)}}:
   \begin{cases}
   x_{u(d)}=\alpha_{yu(d)} \frac{\Delta y_{u(d)}}{\tan \varphi}            \\
   y_{u(d)}=\alpha_{yu(d)} \Delta y_{u(d)}                                 \\
   z_{u(d)}=\alpha_{yu(d)} \frac{\Delta y_{u(d)}}{\tan \theta \sin \varphi}\\
   \end{cases}
\end{alignat}
\begin{alignat}{2} 
   \underset{\in f_{z}}{M_{u(d)}}:
   \begin{cases}
   x_{u(d)}=\alpha_{zu(d)} \Delta z_{u(d)} \tan \theta \cos \varphi \\
   y_{u(d)}=\alpha_{zu(d)} \Delta z_{u(d)} \tan \theta \sin \varphi \\
   z_{u(d)}=\alpha_{zu(d)} \Delta z_{u(d)},                         \\
   \end{cases}
\end{alignat}
\end{subequations}

\noindent where the factors $(\alpha_{xu(d)},\alpha_{yu(d)},\alpha_{zu(d)})$, are equal to $\pm1$, depending on the octant to which
the direction $\bd{n}$ belongs, as detailed in Tables~\ref{tab:octant_u} and ~\ref{tab:octant_d}.

\begin{table*}[ht]
\caption{Correspondence between octant, directional angles $(\theta,\varphi)$, upwind cell indices, sizes of the upwind cell faces in each direction,
 and $(\alpha_{xu},\alpha_{yu},\alpha_{zu})$ factors.
(See Sections~\ref{subsec:overview_SC} and \ref{subsec:intersec_SC} for the definitions of these quantities.)}
\label{tab:octant_u}
\centering
\begin{tabular}{clccccccccc}
\hline\hline
\rule {0pt} {10pt}
  Octant                &  $  (\theta,\varphi)$ (radians)                       &   Upwind~cell     &    Face $f_{x}$     &    Face $f_{y}$     &    Face $f_{z}$     &  $(\alpha_{x u},\alpha_{y u},\alpha_{z u})$ \\
                        &                                                       &     indices       &  $\Delta x_{u}$     &  $\Delta y_{u}$     &  $\Delta z_{u}$     &                                             \\
  \hline \\[0.0cm]
   1                    &  $[0,\frac{\pi}{2}[   \times [0,\frac{\pi}{2}[     $  &  $(i  ,j,  k)  $  &  $|-\Delta x_{i}|$  &  $|-\Delta y_{j}|$  &  $|-\Delta z_{k}|$  &  $ (-1,-1,-1) $                             \\[0.2cm]
   2                    &  $[0,\frac{\pi}{2}[   \times [\frac{\pi}{2},\pi[   $  &  $(i+1,j,  k)  $  &  $\Delta x_{i+1} $  &  $|-\Delta y_{j}|$  &  $|-\Delta z_{k}|$  &  $ (+1,-1,-1) $                             \\[0.2cm]
   3                    &  $[0,\frac{\pi}{2}[   \times [\pi,\frac{3\pi}{2}[  $  &  $(i+1,j+1,k)  $  &  $\Delta x_{i+1} $  &  $\Delta y_{j+1} $  &  $|-\Delta z_{k}|$  &  $ (+1,+1,-1) $                             \\[0.2cm]
   4                    &  $[0,\frac{\pi}{2}[   \times [\frac{3\pi}{2},2\pi[ $  &  $(i  ,j+1,k)  $  &  $|-\Delta x_{i}|$  &  $\Delta y_{j+1} $  &  $|-\Delta z_{k}|$  &  $ (-1,+1,-1) $                             \\[0.2cm]
   5                    &  $[\frac{\pi}{2},\pi[ \times [0,\frac{\pi}{2}[     $  &  $(i  ,j,k+1)  $  &  $|-\Delta x_{i}|$  &  $|-\Delta y_{j}|$  &  $\Delta z_{k+1} $  &  $ (-1,-1,+1) $                             \\[0.2cm]
   6                    &  $[\frac{\pi}{2},\pi[ \times [\frac{\pi}{2},\pi[   $  &  $(i+1,j,k+1)  $  &  $\Delta x_{i+1} $  &  $|-\Delta y_{j}|$  &  $\Delta z_{k+1} $  &  $ (+1,-1,+1) $                             \\[0.2cm]
   7                    &  $[\frac{\pi}{2},\pi[ \times [\pi,\frac{3\pi}{2}[  $  &  $(i+1,j+1,k+1)$  &  $\Delta x_{i+1} $  &  $\Delta y_{j+1} $  &  $\Delta z_{k+1} $  &  $ (+1,+1,+1) $                             \\[0.2cm]
   8                    &  $[\frac{\pi}{2},\pi[ \times [\frac{3\pi}{2},2\pi[ $  &  $(i  ,j+1,k+1)$  &  $|-\Delta x_{i}|$  &  $\Delta y_{j+1} $  &  $\Delta z_{k+1} $  &  $ (-1,+1,+1) $                             \\[0.2cm]
  \hline
\end{tabular}
\tablefoot{We intentionally use the absolute value notations above, in order to emphasize the coordinates, in the local coordinate
           system, of the upwind cell faces that may be intersected by a short-characteristic. For example,
           for octant~1, the $x$-coordinate of $f_{x}$ is $x=-\Delta x_{i}$.}
\end{table*}

\begin{table*}[ht]
\small
\begin{center}
\caption{Same as Table~\ref{tab:octant_u}, but for the downwind cell faces.} \label{tab:octant_d}
\begin{tabular}{clccccccccc}
\hline\hline
\rule {0pt} {10pt}
  Octant                &  $  (\theta,\varphi)$ (radians)                       &   Downwind~cell   &    Face $f_{x}$     &    Face $f_{y}$     &    Face $f_{z}$     &  $(\alpha_{z d},\alpha_{x d},\alpha_{y d})$ \\
                        &                                                       &    indices        &  $\Delta x_{d}$     &  $\Delta y_{d}$     &  $\Delta z_{d}$     &                                             \\
  \hline \\[0.0cm]
   1                    &  $[0,\frac{\pi}{2}[   \times [0,\frac{\pi}{2}[     $  &  $(i+1,j+1,k+1)$  &  $\Delta x_{i+1} $  &  $\Delta y_{j+1} $  &  $\Delta z_{k+1} $  &  $ (+1,+1,+1) $                             \\[0.2cm]
   2                    &  $[0,\frac{\pi}{2}[   \times [\frac{\pi}{2},\pi[   $  &  $(i  ,j+1,k+1)$  &  $|-\Delta x_{i}|$  &  $\Delta y_{j+1} $  &  $\Delta z_{k+1} $  &  $ (-1,+1,+1) $                             \\[0.2cm]
   3                    &  $[0,\frac{\pi}{2}[   \times [\pi,\frac{3\pi}{2}[  $  &  $(i  ,j,  k+1)$  &  $|-\Delta x_{i}|$  &  $|-\Delta y_{j}|$  &  $\Delta z_{k+1} $  &  $ (-1,-1,+1) $                             \\[0.2cm]
   4                    &  $[0,\frac{\pi}{2}[   \times [\frac{3\pi}{2},2\pi[ $  &  $(i+1,j,  k+1)$  &  $\Delta x_{i+1} $  &  $|-\Delta y_{j}|$  &  $\Delta z_{k+1} $  &  $ (+1,-1,+1) $                             \\[0.2cm]
   5                    &  $[\frac{\pi}{2},\pi[ \times [0,\frac{\pi}{2}[     $  &  $(i+1,j+1,k)  $  &  $\Delta x_{i+1} $  &  $\Delta y_{j+1} $  &  $|-\Delta z_{k}|$  &  $ (+1,+1,-1) $                             \\[0.2cm]
   6                    &  $[\frac{\pi}{2},\pi[ \times [\frac{\pi}{2},\pi[   $  &  $(i  ,j+1,k)  $  &  $|-\Delta x_{i}|$  &  $\Delta y_{j+1} $  &  $|-\Delta z_{k}|$  &  $ (-1,+1,-1) $                             \\[0.2cm]
   7                    &  $[\frac{\pi}{2},\pi[ \times [\pi,\frac{3\pi}{2}[  $  &  $(i  ,j,  k)  $  &  $|-\Delta x_{i}|$  &  $|-\Delta y_{j}|$  &  $|-\Delta z_{k}|$  &  $ (-1,-1,-1) $                             \\[0.2cm]
   8                    &  $[\frac{\pi}{2},\pi[ \times [\frac{3\pi}{2},2\pi[ $  &  $(i+1,j,  k)  $  &  $\Delta x_{i+1} $  &  $|-\Delta y_{j}|$  &  $|-\Delta z_{k}|$  &  $ (+1,-1,-1) $                             \\[0.2cm]
  \hline
\end{tabular}
\end{center}
\end{table*}

\section{Piecewise Cubic, Locally Monotonic, Interpolation}   \label{sec:pcw_cubic_mon_intp}

\subsection{The Advantages of a Piecewise, Locally Monotonic, Interpolant}  \label{subsec:interp_interest}
As mentioned in Section~\ref{subsec:overview_SC}, solving the integral form of the radiative transfer equation with the short-characteristics 
method requires to define laws of variation of physical quantities within faces of the grid and along the short-characteristics.
The linear law is the simplest and fastest method. However, it leads to large numerical diffusion. This means that a sharp beam is significantly
dispersed as it propagates throughout a grid, which was discussed in detail by several authors
\citep{Kunasz_and_Auer_1988,Hayek_et_al_2010,Davis_et_al_2012}. Moreover, a second or higher order polynomial is mandatory, in order to recover
the diffusion approximation at large optical depth.
A parabolic law, however, generates overshoots that may be negative and thus
unphysical. This point was discussed in detail by \citet{Kunasz_and_Auer_1988} and \citet{Auer_and_Paletou_1994}.
An efficient method to overcome this difficulty is to use in each direction ($x$, $y$, $z$ and the short-characteristic's direction) a piecewise
continuous, locally monotonic, interpolation function, as proposed by \citet{Auer_2003}. The monotonicity suppresses any spurious extrema. Hereafter,
we describe the method that we have adopted.

\subsection{Cubic Hermite Polynomial and Weighted Harmonic Mean Node Derivative}  \label{subsec:Hermite_and_derivative}
Let us consider a physical quantity that depends on only one variable, $w(x)$, and which is specified only in a set of $m$ discrete positions or nodes
$(x_{i})_{i=1,n}$, with $0 \leqslant x_{1}<x_{2}<...<x_{n}$. 
In other words,  $w_{i}=w(x_{i})$ for $i=1,2,...,n$ are given.
Our objective is to find an interpolation function that goes through all the data points $(x_{i},w_{i})$ and that preserves the monotonicity of the
data. A solution to this problem is provided by a piecewise cubic Hermite polynomial coupled with a weighted harmonic mean defining the
derivative at each node, $w'_{i}=w'(x_{i})$. We detail this solution hereafter.

\begin{figure} 
\plotone{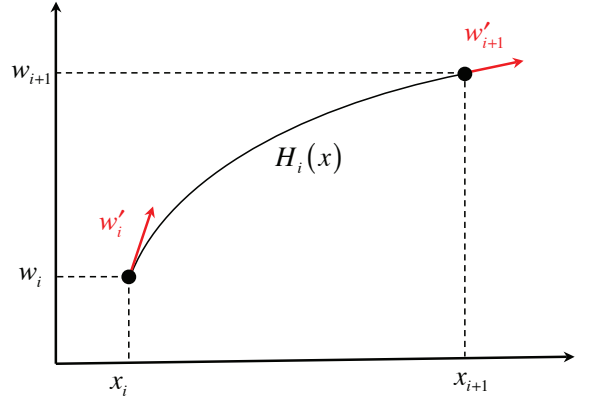}
\caption{Piecewise interpolation. $w(x)$ is the function to be interpolated between the end~points (or nodes) $x_{i}$ and $x_{i+1}$.
We assume that we know $w$ and its derivatives at the end~points: $w_{i}=w(x_{i})$, $w_{i+1}=w(x_{i+1})$, $w'_{i}=w'(x_{i})$, $w'_{i+1}=w'(x_{i+1})$.
The local interpolation function, $H_{i}(x)$, is a cubic Hermite polynomial. It is a third degree polynomial defined between $x_{i}$ and $x_{i+1}$,
from the four relations: $H\left(x_{i}\right)=w_{i}$, $H\left(x_{i+1}\right)=w_{i+1}$, $H'\left(x_{i}\right)=w'_{i}$,
$H'\left(x_{i+1}\right)=w'_{i+1}$. The arrows indicate the slopes of $H(x)$ at the end~points, or, equivalently, the derivatives $w'_{i}$,
$w'_{i+1}$. See Section~\ref{subsec:Hermite_and_derivative} for detailed explanations.}
\label{fig:local_Hermite}
\end{figure}

Among the nodes above, let us choose two consecutive ones, $x_{i}$ and $x_{i+1}$ (see Figure~\ref{fig:local_Hermite}). Within the interval
$[x_{i},x_{i+1}]$, a cubic Hermite polynomial is defined as (see equation~(2) of \citealt{Auer_2003}):
\begin{equation}  \label{eq:Hermite_polynomial}
  \begin{split}
     H_{i}(x) = \left(1-3q_{i}^{2}+2q_{i}^{3}\right) w_{i}                                \\
              + \left(3q_{i}^{2}-2q_{i}^{3}\right) w_{i+1}                                \\
              + \left(q_{i}^{3}-2q_{i}^{2}+q_{i}\right) \left(x_{i+1}-x_{i}\right) w'_{i} \\
	      + \left(q_{i}^{3}-q_{i}^{2}\right)    \left(x_{i+1}-x_{i}\right) w'_{i+1},
  \end{split}
\end{equation}
where
\begin{equation}  \label{eq:q_def}
     q_{i}(x)=\frac{x-x_{i}}{x_{i+1}-x_{i}}, \qquad  0 \leqslant q_{i}(x) \leqslant 1
\end{equation}

By definition, the Hermite polynomial matches the interpolated function $w(x)$ and its derivatives at both ends of the interval:
\begin{subequations}  \label{eq:Hermite_end_values_derivatives}
      \begin{equation}  \label{eq:Hermite_end_values}
            \begin{cases}
               H_{i}\left(x_{i}\right)   = w_{i} \\
               H_{i}\left(x_{i+1}\right) = w_{i+1}
            \end{cases}                   
      \end{equation}
      \begin{equation}  \label{eq:Hermite_end_derivatives}
            \begin{cases}
               H'_{i}\left(x_{i}\right)   = w'_{i} \\
               H'_{i}\left(x_{i+1}\right) = w'_{i+1}
            \end{cases}
      \end{equation}
\end{subequations}

The derivative at an inner node, $(w'_{i})_{i=2,n-1}$, is defined by the weighted harmonic mean suggested by \citet{Brodlie_1980} and
\citet{Fritsch_and_Butland_1984}, and brought to the attention of the astrophysics community by \citet{Auer_2003}:
\begin{subequations}  \label{eq:inner_node_derivative}
      \begin{equation}  \label{eq:derivative}
         w'_{i} =
         \left\{
         \begin{alignedat}{2}
            &\frac{f'_{i-1}f'_{i}}{\left(1-\alpha_{i}\right)f'_{i-1}+\alpha_{i}f'_{i}} & \quad \text{if $f'_{i-1}f'_{i} > 0$,} \\
            &0                                                                         & \quad \text{if $f'_{i-1}f'_{i} \leqslant 0$,}
         \end{alignedat}
         \right.
      \end{equation}
      \text{where} 
      \begin{equation}  \label{eq:notations_fprime}
            f'_{i} = \frac{w_{i+1}-w_{i}}{x_{i+1}-x_{i}}
      \end{equation}
      \begin{equation}  \label{eq:notations_alpha2}
         \alpha_{i} = \frac{1}{3}\left(1+\frac{x_{i+1}-x_{i}}{x_{i+1}-x_{i-1}}\right)
      \end{equation}
\end{subequations}

The derivatives at the outer nodes, $w'_{1}$ and $w'_{n}$, are defined by the local slopes:
\begin{subequations}  \label{eq:outer_node_derivative}
      \begin{equation}  \label{eq:wprime_1}
         w'_{1} =  \frac{w_{2}-w_{1}}{x_{2}-x_{1}}
      \end{equation}
      \begin{equation}  \label{eq:wprime_n}
         w'_{n} =  \frac{w_{n}-w_{n-1}}{x_{n}-x_{n-1}}
      \end{equation}
\end{subequations}

Adopting this definition of the node derivatives, each local cubic Hermite interpolant $(H_{i}(x))_{i=1,n-1}$ is monotonic in the interpolation
interval $[x_{i},x_{i+1}]$. More precisely, the sign of $H_{i}(x)$, for $x \in ]x_{i},x_{i+1}[$, is equal to the sign of $f'_{i}$.
Therefore, spurious extrema never occur, which guarantees the positivity of the interpolated physical quantities.
Let us note $H(x)$ the piecewise interpolation function defined over the whole interval $[x_{1},x_{n}]=\bigcup_{i=1}^{n-1} [x_{i},x_{i+1}]$, so that
its restriction to each interval $[x_{i},x_{i+1}]$ is equal the local function $H_{i}(x)$. $H(x)$ is continuous. Moreover, it is easy to demonstrate
that its derivative $H'(x)$ is also continuous throughout the whole interval $[x_{1},x_{n}]$, specifically at each inner node $(x_{i})_{i=2,n-1}$.

In addition to the Hermite interpolation, \citet{Auer_2003} suggests a second possibility: using a B\'{e}zier polynomial as a local interpolant. The
latter is quite close to a Hermite polynomial. Both functions match the local end~point values $w_{i}$, $w_{i+1}$. However, while the Hermite
polynomial matches the end~point derivatives $w'_{i}$, $w'_{i+1}$, the B\'{e}zier polynomial does not do it necessarily. The reason is as follows.
Let us note $B_{i}(x)$ the local B\'{e}zier polynomial defined in $[x_{i},x_{i+1}]$. $B_{i}(x)$, and, therefore, its derivative $B'_{i}(x)$, depend on
free parameters, called ``control values'' \citep{Auer_2003}. These values can be adjusted in order to control the variation of $B_{i}(x)$ in the
vicinity of the end~points. In particular, they can be restrained within a minimum value and a maximum value (see equation~(11) of
\citealt{Auer_2003}), in order to guarantee the monotonicity of $B_{i}(x)$. In this case, the continuity of the derivative $B'(x)$ of the piecewise
interpolant $B(x)$, which is defined over $[x_{1},x_{n}]$ similarly to $H(x)$, is not necessarily verified at the nodes. In short, whatever the
definition of the node derivatives of $w(x)$, $(w'_{i})_{i=1,n}$, a piecewise Hermite interpolation function, $H(x)$, guarantees that $H'(x)$ matches
these node derivatives and that $H'(x)$ is continuous at the nodes. On the other hand, a piecewise B\'{e}zier interpolation
function $B(x)$, along with an adequate choice of the control values, guarantees to be locally monotonic, but $B'(x)$ does not necessarily match the
node derivatives and is not necessarily continuous at the nodes. Now, if we adopt very specific control values (see equation~(9) of
\citealt{Auer_2003} in the cubic case), we can force the Bezier polynomial to be identical to the Hermite polynomial.
And, as we discussed in the paragraph above, with an adequate choice of the definition of the nodes derivatives $(w'_{i})_{i=1,n}$
(Equations~\ref{eq:inner_node_derivative} and \ref{eq:outer_node_derivative}), a piecewise Hermite interpolation function can be made locally
monotonic.

Finally, let us mention another possible definition of the node derivatives $w'_{i}$, that ensures the monotonicity of a Hermite polynomial.
\citet{Steffen_1990} suggests to calculate the slope at inner node $(x_{i})_{i=2,n-1}$ from the unique parabola passing through the points
$(x_{i-1},w_{i-1}; x_{i},w_{i}; x_{i+1},w_{i+1})$, if this parabola is monotonic in $[x_{i-1},x_{i+1}]$. If not, the node derivative is defined as
the common slope shared by two parabolas, which are locally monotonic in $[x_{i-1},x_{i}]$ and $[x_{i},x_{i+1}]$. Specific definitions are proposed for
the outer nodes derivatives $w'_{1}$ and $w'_{n}$. Although this possibility suggested by \citet{Steffen_1990} satisfies our requirements of 
monotonicity, we did not implement it in IRIS, because the calculation of such a derivative, for a given node, uses a larger cpu time than the
calculation with the weighted harmonic mean formula (Equation~(\ref{eq:derivative}) in this paper may be compared to equation~(11) of
\citealt{Steffen_1990}).

\section{Implementation}   \label{sec:propagation_ray}

\subsection{Interpolations in Cell Faces, Ghost Nodes}  \label{subsec:intp_cell_faces}
For each direction of propagation of the radiation, we apply the 3D~short-characteristics method, which consists in solving the integral form of the
RTE, Equation~(\ref{eq:RTE3}), throughout the 3D~Cartesian grid points. We repeat this procedure for all the directions specified over a solid angle.
The method that we employ for propagating the radiation gives a privileged role to $z$-axis, in the sense that we are sweeping the 3D~grid gradually,
$z$-plane by $z$-plane, in the direction of increasing (or decreasing) $z$. 
In an astrophysical context, $z$-axis should represent a global direction 
of the energy transport in an object; that is the direction from the interior to the outer layers of a stellar or a planetary atmosphere, 
the vertical distance from an accretion disk plane, or the direction of an accretion column in a young stellar object.

Boundary conditions must be defined in appropriate planes, which depend on the direction of propagation of the radiation. Assuming that the grid is
defined by $N_{x}$ cells in $x$-direction, $N_{y}$ cells in $y$-direction, and $N_{z}$ cells in $z$-direction, the positions of the grid nodes in
each direction may be noted as $(x_{0},x_{1}, ... ,x_{N_{x}})$, $(y_{0},y_{1}, ... ,y_{N_{y}})$, and  $(z_{0},z_{1}, ..., z_{N_{z}})$.
The sizes of the cells in each direction are then noted $(\Delta x_{1}, ... ,\Delta x_{N_{x}})$, $(\Delta y_{1}, ... ,\Delta y_{N_{y}})$,
and $(\Delta z_{1}, ... ,\Delta z_{N_{z}})$, and they are defined by:
\begin{equation}  \label{eq:cell_sizes}
      \begin{cases}
         \Delta x_{i} = x_{i}-x_{i-1} \\
         \Delta y_{j} = y_{j}-y_{j-1} \\
         \Delta z_{k} = z_{k}-y_{k-1}
      \end{cases}                   
\end{equation}

In agreement with our notations in Fig.~\ref{fig:SC_global}, a propagation in the first octant corresponds to directional angles
$(\theta,\varphi) \in [0,\pi/2[ \times [0,\pi/2[$, and, therefore, to a propagation along increasing $x$,
increasing $y$, and  increasing $z$. Consequently, the boundary specific intensities must be known at the bottom frontiers of the 3D~Cartesian grid,
defined by the following planes $x=x_{0}$, $y=y_{0}$, $z=z_{0}$.

As explained in Section~\ref{subsec:overview_SC}, calculating the formal solution of the monochromatic RTE in its integral form, at a given grid point
and for a given direction of propagation of the radiation, requires the determination of $(I_{u},S_{u},\chi_{u})$ at the intersection of the
short-characteristic with the upwind face, and $(S_{d},\chi_{d})$ at the intersection of the short-characteristic with the downwind face.

\begin{figure} 
{\centering \leavevmode \includegraphics[width=1.00\columnwidth]{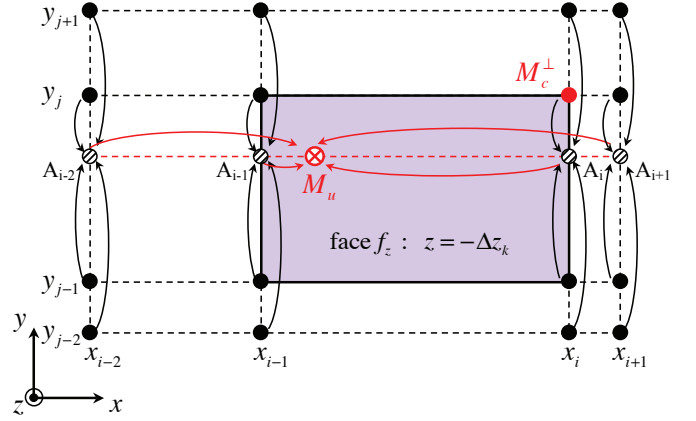}}
{\centering \leavevmode \includegraphics[width=1.00\columnwidth]{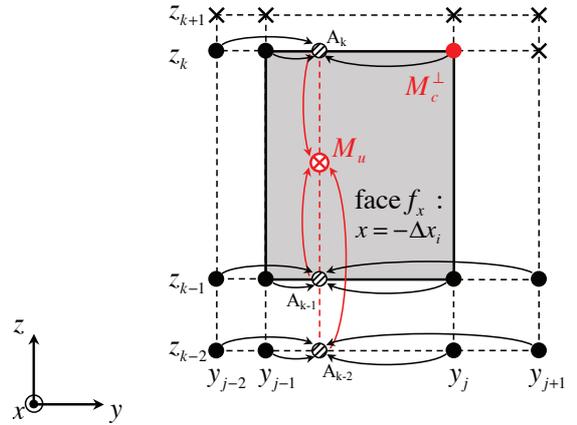}}
\caption{Examples of cubic monotonic interpolations in cell faces intersected by a short-characteristic, for a ray propagating in the first octant,
i.e., along increasing $x$, $y$, and $z$. Two configurations are possible. The first one is displayed in the top panel, where all the quantities are
known in the grid points neighboring the cell face. This situation always occurs for the upwind and downwind source functions and opacities
$(S_{u},\chi_{u})$, $(S_{d},\chi_{d})$. But, for the upwind specific intensity $I_{u}$, it occurs only when horizontal faces 
$f_{z}: z=\pm\Delta z_{k}$ or boundary faces are intersected. The second configuration is displayed in the bottom panel. It impacts the determination
of $I_{u}$ when vertical non-boundary faces, $f_{x}: x=\pm\Delta x_{i}$ or $f_{y}: y=\pm\Delta y_{j}$, are intersected. In this case, the upwind
specific intensity is not known in all the grid points neighboring the cell face. See Section~\ref{subsec:intp_cell_faces} for detailed explanations.}
\label{fig:interp_faces}
\end{figure}

The opacities $\chi$ are known at each point of the three-dimensional grid. And, by definition of a formal solution, the source function $S$
is known at each grid point. Then, $S_{u}$, $\chi_{u}$, (respectively $S_{d}$, $\chi_{d}$) are determined by successive cubic monotonic interpolations
in each of the two directions of a given upwind (respectively downwind) horizontal of vertical face. The top panel of Figure~\ref{fig:interp_faces}
provides an illustration of a sequence of interpolations in a horizontal upwind cell face. This example corresponds to the case shown in the top panel
of Fig.~\ref{fig:SC_local}. The upwind face is identified by indices $i$ and $j$ and belongs to the plane defined by $f_{z}: z=-\Delta z_{k}$ in the
local coordinate system (see Section~\ref{subsec:intersec_SC}). The ray propagates in the first octant, i.e., along increasing $x$, $y$, and $z$.
The black dots indicate the grid points. $M_{u}$ is the upwind end~point. $M_{c}^{\perp}$ is the
projection of the current point $M_{c}$ on $f_{z}$, and coincides with a grid point. Quantities defined in the grid points are interpolated in
$y$-direction at the $y_{u}$ position of $M_{u}$. Then, the interpolated values, in positions $\rm A_{i-2}$, $\rm A_{i-1}$, $\rm A_{i}$, 
$\rm A_{i+1}$, are themselves interpolated in $x$-direction at the $x_{u}$ position of $M_{u}$, which provides the upwind quantity $S_{u}$,
$\chi_{u}$. The curved arrows indicate the grid points which contribute to the interpolations. Four points are necessary to accomplish
each interpolation (see Section~\ref{subsec:Hermite_and_derivative} along with Equations~(\ref{eq:Hermite_polynomial}) and
(\ref{eq:inner_node_derivative})).

Unlike opacities and source functions, the upwind specific intensity is not always known in all the grid points neighboring the upwind cell face.
The sweeping procedure throughout the 3D~Cartesian grid is made $z$-plane by $z$-plane, as described above in the first paragraph of this section.
Therefore, for each new $z$-plane to be processed, the specific intensity is known at each grid point of the upwind $z$-plane. Accordingly, the
procedure described above for $S$ and $\chi$ works if $M_{u}$ belongs to a horizontal face.
However, the specific intensity is not known at all grid points of a preceding vertical plane ($f_{x}: x=\pm\Delta x_{i}$, or
$f_{y}: y=\pm\Delta y_{j}$, in the local coordinate system), except if it is a boundary plane. The bottom panel of Fig.~\ref{fig:interp_faces}
provides an example of a sequence of interpolations of the specific intensity in a vertical upwind cell face. In this case, $z=z_{k}$ is the current
$z$-plane, $z=z_{k-1}$ is the upwind one. $M_{u}$ belongs to the vertical face identified by indices $j$ and $k$, and by $f_{x}: x=-\Delta x_{i}$. We
still consider a ray propagating in the first octant (increasing $x$, $y$, $z$). 
The specific intensity is fully known in the preceding $z=z_{k-2}$ and $z=z_{k-1}$ planes, where we can perform
interpolations in $y$-direction at the $y_{u}$ position of $M_{u}$. This provides the values of the specific intensity in points
$\rm A_{k-2}$ and $\rm A_{k-1}$. The crosses indicate the grid points in which the specific intensity is not known. Since the sweeping follows the
direction of propagation of the ray, it is performed along increasing $y$, in the $z=z_{k}$ plane. Therefore, the specific
intensity is known in the three consecutive grid points ($(y_{j-2},z_{k})$, $(y_{j-1},z_{k})$, $(y_{j},z_{k})$), but not in $(y_{j+1},z_{k})$. It is
also not known in the $z=z_{k+1}$ plane (except in $y=y_{j-2}$ if this $y$-plane is a boundary). Consequently, we can determine the specific
intensity in point $\rm A_{k}$ in $z=z_{k}$ plane, with a cubic Hermite interpolation in $y$-direction, using the values in 
($(y_{j-2},z_{k})$, $(y_{j-1},z_{k})$, $(y_{j},z_{k})$). 
Finally, the interpolated values in $\rm A_{k-2}$, $\rm A_{k-1}$, $\rm A_{k}$ are themselves interpolated in $z$-direction at the $z_{u}$ position of
$M_{u}$, which provides the upwind specific intensity $I_{u}$.
For the last two interpolations, since we know the intensities in only three grid points, the derivatives of the cubic Hermite polynomials in
the edge points, $M_{c}^{\perp}$ in $y$-direction and $\rm A_{k}$ in $z$-direction, are provided by the
local slopes as defined in Equations~(\ref{eq:outer_node_derivative}).

We have added at each boundary of each of the three $x$, $y$, $z$ dimensions, two layers of ghost nodes that embed the initial 3D~Cartesian grid.
These ghost nodes are necessary if we want to apply the above interpolation procedure uniformly to all the grid points, which spares us from testing
whether the grid points are close to the boundaries or not. In $x$-direction, for example, and for
a propagation in increasing $x$, ghost nodes are required for determining the upwind quantities at $x_{1}$, and for determining the downwind
quantities at $x_{N_{x}-1}$ and $x_{N_{x}}$. The ghost nodes $(x_{-2},x_{-1})$ and $(x_{N_{x+1}},x_{N_{x+2}})$ are added.
Their positions per se do not matter, as we define in these nodes positive linear extrapolations of the state parameters.
Then, it is easy to show that we can still use the cubic Hermite polynomial interpolants in $x_{1}$, $x_{N_{x}-1}$ and $x_{N_{x}}$,
and that the derivatives in these nodes, even though calculated with the weighted harmonic mean, are equal to the local slopes defined by
Equations~(\ref{eq:outer_node_derivative}).

\subsection{The Discretized Integrated RTE along a Short-Characteristic}  \label{subsec:sc_discretized}
In the preceding subsection, we have described our procedure to determine the upwind quantities $(I_{u},S_{u},\chi_{u})$, and the downwind quantities
$(S_{d},\chi_{d})$. Once we know them, we are in a position to calculate the integral form of the RTE, Equation~(\ref{eq:RTE3}).

To do so, we define laws of variation, along the short-characteristics, of the source function and the opacity. We use a monotonic cubic Hermite
polynomial between the upwind endpoint $M_{u}$ and the current grid point $M_{c}$ (see Fig.~\ref{fig:SC_global}). Since we know the values of $S$
and $\chi$ in three points, $M_{u}$, $M_{c}$, and the downwind endpoint $M_{d}$, the derivatives of these quantities at $M_{u}$,
$\left. \frac{dS}{d\tau} \right|_{u}$ and $\left. \frac{d\chi}{ds} \right|_{u}$, are provided by the
slopes defined by Equation~(\ref{eq:wprime_1}), while the derivatives at $M_{c}$, $\left. \frac{dS}{d\tau} \right|_{c}$ and
$\left. \frac{d\chi}{ds} \right|_{c}$, are calculated with the weighted harmonic mean,
Equation~(\ref{eq:inner_node_derivative}). Then, the RTE (Equation~(\ref{eq:RTE3})) can be integrated analytically. We obtain the following
expression:
\begin{subequations}  \label{eq:RTE4}
      \begin{equation}  \label{eq:Ic}
         I_{c} = I_{u} e^{-\tau_{uc}}  +  \alpha S_{u} +  \beta S_{c} +  \alpha' \left. \frac{dS}{d\tau} \right|_{u} +  \beta'  \left. \frac{dS}{d\tau} \right|_{c}
      \end{equation}
      \text{where} 
      \begin{equation}  \label{eq:alpha}
            \alpha = \frac{1}{\tau_{uc}^{3}} \{ 6\left(\tau_{uc}-2\right) + \lbrack -\tau_{uc}^{3} + 6\left(\tau_{uc}+2\right)  \rbrack e^{-\tau_{uc}} \}
      \end{equation}
      \begin{equation}  \label{eq:beta}
            \beta  = \frac{1}{\tau_{uc}^{3}} \{\lbrack \tau_{uc}^{3} - 6\left(\tau_{uc}-2\right) \rbrack - 6\left(\tau_{uc}+2\right)    e^{-\tau_{uc}} \}
      \end{equation}
      \begin{equation}  \label{eq:alpha_prime}
            \alpha' = \frac{1}{\tau_{uc}^{2}} \{ \left(2\tau_{uc}-6\right) + \left(\tau_{uc}^{2}+4\tau_{uc}+6\right)    e^{-\tau_{uc}} \}
      \end{equation}
      \begin{equation}  \label{eq:beta_prime}
            \beta' = \frac{1}{\tau_{uc}^{2}} \{ \left(-\tau_{uc}^{2}+4\tau_{uc}-6\right) + \left(2\tau_{uc}+6\right)    e^{-\tau_{uc}} \}
      \end{equation}
\end{subequations}

The optical depth $\tau_{uc}$ is calculated by integrating analytically Equation~(\ref{eq:optical_depth}) from $M_{u}$ to $M_{c}$. We obtain:
\begin{equation}  \label{eq:tau_uc}
      \tau_{uc} = \frac{1}{2} s_{uc} \left( \chi_{u}+\chi_{c} \right) + \frac{1}{2} s_{uc}^{2} \left( \left. \frac{d\chi}{ds} \right|_{u} - \left. \frac{d\chi}{ds} \right|_{c} \right)
\end{equation}
where $s_{uc}$ is the path length from $M_{u}$ to $M_{c}$.

When $\tau_{uc}$ is small enough ($\lesssim 5\times 10^{-2}$), we use Taylor expansions up to the third order of Equations~(\ref{eq:RTE4}).
The bottom line is that the RTE can always be calculated, including the two asymptotic limits: the optically thin or even transparent path for which
$\tau_{uc}$ equals zero, and, therefore, $I_{c} \longrightarrow I_{u}$, and the optically thick or even opaque path for which $\tau_{uc}$ tends to
infinity, and, therefore, $I_{c} \longrightarrow S_{c}$.

\subsection{The Velocity Gradient Effect (Doppler shift)}  \label{subsec:velocity_effect}
We consider a medium with a non-zero macroscopic velocity with respect to the observer's frame. In this case, the velocity gradient between two
positions in the medium causes a Doppler shift of photon frequencies, and therefore a shift
of the corresponding spectral lines.
This effect must be properly taken into account in an implementation of the short-characteristics method. 
If the velocity difference between the current point $M_{c}$ and the upwind end~point $M_{u}$ (see Fig.~\ref{fig:SC_global}) is large enough, then a
simple interpolation scheme may underestimate or overestimate opacities.

To illustrate this point, let us consider a single emission line and a given frequency at which we solve the RTE. The following situation may happen.
At $M_{u}$, the frequency is located at a wing of the line, quite far from the center of the line, such that the absorption coefficient is very small
compared to the line center value. And, at $M_{c}$, the line is shifted, such that this same frequency is at the opposite side of the line, in the
other wing, far from the center. The law of variation then infers a very weak opacity for this line throughout the path from $M_{u}$ to $M_{c}$.
In fact, the line goes through its maximum value at this frequency, somewhere between $M_{u}$ and $M_{c}$, and this is not taken
into account. This issue was pointed out by \citet{van_Noort_et_al_2002} who emphasized the fact that the opacity along a short-characteristic may be
underestimated by several orders of magnitude. In addition to that example, we provide here a case of overestimation. Let us consider now two
lines that are quite distant from each other. We may face the following situation. At $M_{u}$, the calculation frequency can be located close to the
center of one line. And, at $M_{c}$, due to the shift of the lines, this frequency can be close to the maximum of the second line. Consequently, a
variation law defined between $M_{u}$ and $M_{c}$ overestimates the opacity if the two line centers are distant enough, such that we miss the trough
between them.

A common solution of these problems is provided by a subgriding (e.g., \citet{van_Noort_et_al_2002}).
Let $\bd{\rm{v}}$ be the macroscopic velocity  in the observer's frame, and
$\hat{\rm{v}}=\bd{\rm{v}} \cdot \bd n$ its projection on the direction of photon propagation $\bd n$. 
The idea of subgriding is to subdivide a short-characteristic into a set of subintervals, so that the difference in
$\hat{\rm{v}}$ between two points of this subdivision remains small compared to the thermal velocity. In IRIS, we follow this procedure.
Specifically, we introduce a dimensionless parameter factor $\epsilon_{\sm D}$ that provides to the user a control over the subdivision.
Let us identify with indices $l$ and $l+1$ two points of the subdivision of a given short-characteristic. Let us consider a single line. We note
$\nu_{0}$ the transition frequency (emission or absorption of a photon) in the particle's frame (or comoving frame). $\nu_{0}$ coincides with the center of the line in
the fluid rest frame where ${\bd{\rm{v}}} = {\bd 0}$, but this frequency is shifted to the value $\nu_{\rm ctr}$ in the observer's frame where
${\bd{\rm{v}}} \ne {\bd 0}$, as follows,
\begin{equation}  \label{eq:nu_center}
      \nu_{\rm ctr} = \nu_{0} \left( 1+ \tfrac{\hat{\rm{v}}}{c} \right)
\end{equation}
where we consider a Doppler shift of frequencies to first order in $\hat{\rm{v}}/c$, since we assume the velocities to be non relativistic (see Section~\ref{sec:RTE}).
$\nu_{\rm ctr}$ is the center of the line in the observer's frame.
The shift of $\nu_{\rm ctr}$ between position $l$ and position $l+1$ is given by
\begin{equation}  \label{eq:Doppler_shift}
      \nu_{\rm ctr}(l+1)-\nu_{\rm ctr}(l) =  \nu_{0} \tfrac{\hat{\rm{v}}(l+1)-\hat{\rm{v}}(l)}{c}
\end{equation}
The Doppler width associated with this line, in a given position in the medium, is
\begin{equation}  \label{eq:Doppler_width}
      \Delta \nu_{\sm D} = \nu_{0} \frac{V_{\rm th}}{c}
\end{equation} 
with $c$ being the speed of light, and $V_{\rm th}$ the thermal velocity.
The latter is defined as 
\begin{equation}  \label{eq:thermal_velocity}
V_{\rm th}=\left( \frac{2\,k\,T}{m}\right)^{1/2}
\end{equation} 
where $k$ is the Boltzmann constant, $m$ the mass of the particle.
$\epsilon_{\sm D}$ is a control parameter defined such that the Doppler shift of the center of the line between position $l$ and position $l+1$ is bounded as follows:
\begin{equation}  \label{eq:delta_nu_shift}
      |\nu_{\rm ctr}(l+1)-\nu_{\rm ctr}(l)| \leqslant \epsilon_{\sm D} \, \frac{\Delta \nu_{\sm D}(l) + \Delta \nu_{\sm D}(l+1)}{2} 
\end{equation}
The quantity $\frac{1}{2} \lbrack \Delta \nu_{\sm D}(l) + \Delta \nu_{\sm D}(l+1) \rbrack$ is an average of the local Doppler width between
its value in $l$ position and its value in $l+1$ position.
It is easy to demonstrate that this inequality is independent of the transition frequency in the particle's frame $\nu_{0}$, and that 
it is equivalent to the following condition on the gradients of the projected velocities:
\begin{equation}  \label{eq:delta_V_shift}
      |\hat{\rm{v}}(l+1)-\hat{\rm{v}}(l)| \leqslant \epsilon_{\sm D} \,  \frac{V_{\rm th}(l) + V_{\rm th}(l+1)}{2}.
\end{equation}
$\epsilon_{\sm D}$ is a tunable parameter, whose value depends on the
required accuracy of the solution (ideally between 1/3 and 1). The smaller it is, the higher accuracy is achieved; however 
at the expense of increased computer demands.

\section{Radiation Moments and Angular Integration}  \label{sec:moments_angular_integration}

In addition to computing specific intensity, one can calculate important angle-averaged quantities related to radiation moments, such as
the mean intensity $J(\bd{r},\nu,t)$, the radiation flux vector $\bd F(\bd{r},\nu,t)$,
and the radiation pressure tensor $\bd{\mathsf P}(\bd{r},\nu,t)$. 
These are defined as
\begin{equation}  \label{eq:J}
      J(\bd{r},\nu,t) = \frac{1}{4\pi} \oint_{}^{} I(\bd{r},\bd{n},\nu,t)\, d\Omega,
\end{equation}
\begin{equation}  \label{eq:F_vector}
      \bd F(\bd{r},\nu,t) = \oint_{}^{} I(\bd{r},\bd{n},\nu,t) \, \bd{n} \, d\Omega,
\end{equation}
\begin{equation}  \label{eq:P_dyadic}
      \bd{\mathsf P}(\bd{r},\nu,t) = \frac{1}{c} \oint_{}^{} I(\bd{r},\bd{n},\nu,t) \, \bd{n} \bd{n} \, d\Omega,
\end{equation}
where $d\Omega$ is the elementary solid angle, and the symbol $\oint_{}^{}$ designates an integration over $4\pi$~sr.
The flux vector and the radiation pressure tensor can also be written in component form:
\begin{equation}  \label{eq:F_components}
      F_{i} = \oint_{}^{} I(\bd{r},\bd{n},\nu,t) \, n_{i} \, d\Omega,
\end{equation}
\begin{equation}  \label{eq:P_components}
      P_{ij} = \frac{1}{c} \oint_{}^{} I(\bd{r},\bd{n},\nu,t) \, n_{i} n_{j} \, d\Omega,
\end{equation}
In Cartesian coordinates, the three directions (1,2,3) are the unit vectors $(\bd{e_{x}},\bd{e_{y}},\bd{e_{z}})$
introduced in Section~\ref{subsec:overview_SC}, and the components of the radiation propagation vector $\bd{n}$ are
\begin{equation}  \label{eq:n_components}
     \left\{
         \begin{alignedat}{4}
            n_{1} &= n_{x} &= \bd{n} \cdot \bd{e_{x}} &= \sin \theta \cos \varphi \\
            n_{2} &= n_{y} &= \bd{n} \cdot \bd{e_{y}} &= \sin \theta \sin \varphi \\
            n_{3} &= n_{z} &= \bd{n} \cdot \bd{e_{z}} &= \cos \theta
         \end{alignedat}
     \right.
\end{equation}
IRIS computes each of the three components of the radiation flux vector, $F_{x}$, $F_{y}$, $F_{z}$.
Since the radiation pressure tensor is symmetric \citep{Mihalas_1978,Mihalas_and_Mihalas_1984}, it is defined by six components, $P_{xx}$, $P_{yy}$,
$P_{zz}$,  $P_{xy}$, $P_{xz}$, $P_{yz}$, all of them being computed by IRIS.

The angular integrations are numerically performed by using quadratures. A quadrature is a set of $M$ discretized direction vectors
$\bd{n_{m}}=(n_{mx},n_{my},n_{mz})$ and the corresponding weights $w_{m}$, which are the numerical representation of, respectively, the direction
vector $\bd{n}$ and the elementary solid angle $d\Omega$. An appropriate quadrature for an integration over the whole $4\pi$~sr range should satisfy
at least the first two of the three following conditions:
\begin{enumerate}\itemsep-0.04in
\vspace{-10pt}
\item symmetry: the directions of the quadrature are invariant after rotation of $90\degr$ around the three coordinate directions
$(\bd{e_{x}},\bd{e_{y}},\bd{e_{z}})$,
\item for an isotropic radiation field, the three moments should be numerically exact.
\item for an isotropic radiation field, the half first moment should be numerically exact.
\end{enumerate}
\vspace{-6pt}
The second condition reads as follows:
\begin{subequations}  \label{eq:moments_isotropic}
      \begin{alignat}{3}
         & \oint_{}^{} \, d\Omega                  &= & \; 4\pi  \; &= & \; \sum_{m=1}^{M} w_{m} ,            \\
         & \oint_{}^{} \bd{n} \, d\Omega           &= & \; \bd{0}\; &= & \; \sum_{m=1}^{M} w_{m} \bd{n_{m}} , \\
         & \oint_{}^{} \bd{n} \bd{n} \, d\Omega \; &= & \; \frac{4\pi}{3} \bd{\mathsf \delta}  &= & \; \sum_{m=1}^{M} w_{m} \bd{n_{m}} \bd{n_{m}},
      \end{alignat}
\end{subequations}
where $\bd{\mathsf \delta}$ is the unit tensor.
The third condition is optional, and is related to the following issue. One may need to calculate a radiative flux over a surface or at a wall. In
this case, the flux is determined by an integration over $2\pi$~sr, instead of $4\pi$~sr: $\bd F_{\rm wall} = \int_{2\pi}^{} I \, \bd{n} \, d\Omega$.
The third condition reads then:
\begin{equation}  \label{eq:half_1st_moment_isotropic}
   \int_{\bd{n} \cdot \bd{e_{i}} > 0}^{} \bd{n} \, d\Omega  = -\int_{\bd{n} \cdot \bd{e_{i}} < 0}^{} \bd{n} \, d\Omega  = \pi \bd{e_{i}} = \sum_{\bd{n_{m}} \cdot \bd{e_{i}} > 0}^{} w_{m} \bd{n_{m}}
\end{equation}
where $\bd{e_{i}}$ stands for $e_{x}$, $e_{y}$, or $e_{z}$, and where we consider surfaces or walls perpendicular to one of these three
directions.

We have implemented in IRIS the quadratures of type A proposed by \citet{Carlson_1963}, and whose construction principle is
summarized by \citet{Bruls_et_al_1999}. They satisfy conditions 1 and 2, but not condition 3. We also have implemented the quadratures built by
\citet{Lathrop_and_Carlson_1965}. The latter provide several sets of quadratures that satisfy conditions 1 and 2, but not condition 3, with
24 up to 288~directions. They also provide several sets of another type of quadratures that satisfy the three conditions above, with
24 up to 80~directions.

We also have implemented the Gaussian quadrature \citep{Press_et_al_1992}. The latter is not appropriate for the calculation of the radiation moments,
because it does not
satisfy any of the three conditions above and leads to serious non physical asymmetries, unless one chooses a large number of directions (typically,
more than 100). However, it is suitable for the integration of the incoming specific intensity over a viewing solid angle, in order to determine the
radiative power per unit surface that receives a detector looking at a given object.

\section{Periodic Boundary Conditions}  \label{sec:PBC}

We consider a three-dimensional medium with an infinite extension in the horizontal plane $\left(x,y\right)$, and a finite extension along the
vertical $z$-axis between its lower boundary $z_{0}$ and its upper boundary $z_{N_{z}}$. We assume that this medium has a double periodicity, one
in $x$-direction, and one in $y$-direction. The boundary conditions are known at
$z_{0}$ and $z_{N_{z}}$. For example, we may consider a non-irradiated
stellar atmosphere with no incoming radiation at the outer surface $z_{N_{z}}$ and a black body radiation at its inner surface $z_{0}$.
The computational grid ranges from $z_{0}$ to $z_{N_{z}}$ in vertical direction, from $x_{0}$ to $x_{N_{x}}$, and from $y_{0}$ to
$y_{N_{y}}$ in the horizontal plane, so that $(x_{N_{x}}-x_{0})$ defines one $x$-period and $(y_{N_{y}}-y_{0})$
defines one $y$-period. Now, the 3D short-characteristics method consists in solving the integral form of the RTE by propagating the rays
throughout a computational domain, from up to three upwind boundary sides in which the specific intensity is assumed to be known, down up to the
three other faces of the domain.
For horizontally periodic media, while the vertical boundary conditions are specified explicitly, the lateral boundary
conditions are  defined implicitly, such that, for any physical quantity $f(\nu,x,y,z)$, we have the following relations:
\begin{subequations}  \label{eq:f_periodic}
\begin{align}
   f(\nu,x_{N_{x}},y,z) = f(\nu,x_{0},y,z), \text{\quad for any } y,z,    \label{eq:f_periodic_x}\\
   f(\nu,x,y_{N_{y}},z) = f(\nu,x,y_{0},z), \text{\quad for any } x,z.    \label{eq:f_periodic_y}
\end{align}
\end{subequations}

\noindent Consequently, when the upwind end~point of a short-characteristic intersects a lateral boundary, we prolong this characteristic, which
becomes a long-characteristic, until it intersects a horizontal face, following the suggestions by \citet{Auer_et_al_1994} and
\citet{Fabiani_Bendicho_2003}. For a given direction of propagation, and for each $z$-layer, this treatment affects only the rows that are adjacent to
the boundary faces.

\begin{figure} 
\plotone{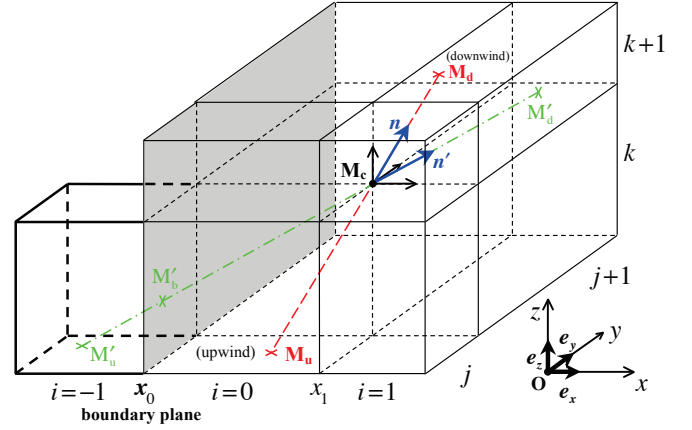}
\caption{Extract of a 3D Cartesian grid. Unlike the general case presented in Fig.~\ref{fig:SC_global}, we consider here the particular case of a
medium with a double horizontal periodicity, in $x$-direction and in $y$-direction. In $x$-direction, one period is defined from $x_{0}$ to
$x_{N_{x}}$. The plane $x=x_{0}$ is a boundary plane. The cell defined by indices $(-1,j,k)$, drawn with thick lines,
is identical to the cell $(N_{x},j,k)$ (not drawn). The upwind specific intensity is known at $M_{u}$ for the short-characteristic $(M_{u},M_{c},M_{d})$, since
$M_{u}$ belongs to a horizontal face. On the other hand, the upwind specific intensity is not known at $M'_{b}$ for the short-characteristic
$(M'_{b},M_{c},M'_{d})$, because $M'_{b}$ belongs to a vertical boundary face. Therefore, the characteristic is prolonged up to the first intersection
with a horizontal face, $M'_{u}$. See Section~\ref{sec:PBC} for more explanations.}
\label{fig:PBC}
\end{figure}

Figure~\ref{fig:PBC} illustrates our point with an example showing a propagation of the radiation in the first octant, i.e., along increasing $x$,
$y$, and $z$. Two characteristics are plotted from the current point $M_{c}$ in the plane $x=x_{1}$, which is adjacent to the boundary plane
$x=x_{0}$. The upwind end~point $M_{u}$ of the short-characteristic $(M_{u},M_{c},M_{d})$ belongs to a horizontal face. Therefore, the specific
intensity can be calculated by interpolations from the values at the neighboring grid points in the upwind horizontal plane. If we consider
the short-characteristic $(M'_{b},M_{c},M'_{d})$, the upwind end~point $M'_{b}$ belongs to the horizontal boundary plane $x=x_{0}$, in which the
specific intensity is not explicitly defined. The short-characteristic is then lengthened, and becomes a long-characteristic, down to its first
intersection, $M'_{u}$, with a horizontal face. The long-characteristic can go through several cells (only one cell in
the example, for the clarity of the figure) beyond the vertical boundary face, before hitting a horizontal face.

When the medium is not periodic, we start the sweeping of a given $z$-plane at the grid point that is the closest to the
vertical boundary planes. For example, if the ray propagates in the first octant, these planes are $x=x_{0}$ and $y=y_{0}$, therefore the first grid
point in which we calculate the specific intensity is $(x_{1},y_{1})$. Then, the specific intensity is calculated progressively in all the grid points
in the $z$-plane along $x$-direction and $y$-direction.
Now, as we consider in this section a medium with a double horizontal periodicity, we can start the sweeping at any grid point in
the $z$-plane, provided that the grid points where the specific intensity is calculated span one $x$-period and one $y$-period.
We mark this starting point as $(x_{i_{0}},y_{j_{0}})$. The indices $i_{0}$ and $j_{0}$
are defined so as to minimize the number of cells that are intersected by the long-characteristics, thus saving computing time. This way,
$i_{0}$ is chosen such that the corresponding upwind cell has the largest size in $x$-direction, $\Delta x_{i_{0}}=\Delta x_{\rm{max}}$. 
Similarly, we choose $j_{0}$ such that $\Delta y_{j_{0}}=\Delta j_{\rm{max}}$. Then, starting from
$(x_{i_{0}},y_{j_{0}})$, we calculate the specific intensity with the long-characteristics method at $y_{j_{0}}$, for the grid points along
$x$-direction that span one $x$-period.
In the same vein, we use the long-characteristics method at $x_{i_{0}}$, for the grid points along $y$-direction
that span one $y$-period. Once these two first rows have been treated, we resume
the short-characteristics method for all the next rows of the current $z$-layer.
Note that here we do not introduce ghost nodes (see Section~\ref{subsec:intp_cell_faces}), since the medium has an infinite extension in 
$x$-direction and $y$-direction.

\begin{figure} 
\plotone{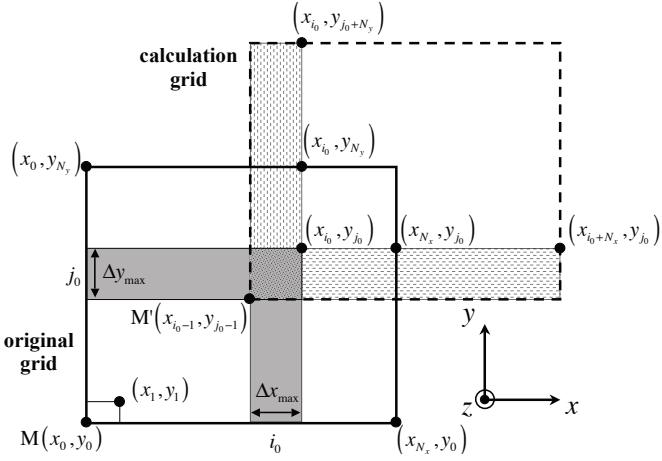}
\caption{Original domain (solid thick lines) and calculation domain (dotted thick lines) at a given $z$-plane, for a medium that is periodic in the
horizontal ($x$,$y$) plane, and for a radiation propagating in the first octant, i.e., along increasing $x$, $y$, and $z$. Both domains span one
period in $x$-direction and one period in $y$-direction. The nodes of the original domain are 
$(x_{0}, x_{1}, ... , x_{N_{x}})$ and $(y_{0}, y_{1}, ... , y_{N_{y}})$. The nodes of the calculation domain, where the specific intensity is
computed, are $(x_{i_{0}}, x_{i_{0}+1}, ... , x_{i_{0}+N_{x}})$ and $(y_{j_{0}}, y_{j_{0}+1}, ... , y_{j_{0}+N_{y}})$.
The indices $i_{0}$ and $j_{0}$ are chosen such that $\Delta x_{i_{0}}=\Delta x_{\rm{max}}$ and $\Delta y_{j_{0}}=\Delta j_{\rm{max}}$.
See text in Section~\ref{sec:PBC} for a discussion.}
\label{fig:PBC_starting_rows}
\end{figure}

The schematic diagram of Figure~\ref{fig:PBC_starting_rows} clarifies our point, for a radiation propagating in the first octant,
therefore along increasing $x$, $y$, and $z$. It shows two computational domains in a given $z$-plane. Both span one $x$-period and one
$y$-period. The original domain is depicted by the rectangle with solid thick lines. The positions of the grid nodes in the horizontal
$z$-plane were introduced in Section~\ref{subsec:intp_cell_faces}, $(x_{0},x_{1}, ... ,x_{N_{x}})$ and $(y_{0},y_{1}, ... ,y_{N_{y}})$. The
vertical boundary planes are $x=x_{0}$ and $y=y_{0}$.
The indices $i_{0}$ and $j_{0}$ are defined above in the preceding paragraph.
The RTE is solved in the computational domain depicted by the rectangle with thick dotted lines. The positions of the grid nodes of this domain are
$(x_{i_{0}}, x_{i_{0}+1}, ... , x_{i_{0}+N_{x}})$ and $(y_{j_{0}}, y_{j_{0}+1}, ... , y_{j_{0}+N_{y}})$, with the following correspondence of the
grid sizes, derived from the double periodicity of the medium:
\begin{equation}  \label{eq:comput_domain}
      \begin{cases}
         \Delta x_{i,\,i\in \llbracket N_{x}+1,N_{x}+i_{0} \rrbracket}=\Delta x_{i',\,i'\in \llbracket 1,i_{0} \rrbracket} \\
         \Delta y_{j,\,j\in \llbracket N_{y}+1,N_{y}+j_{0} \rrbracket}=\Delta y_{j',\,j'\in \llbracket 1,j_{0} \rrbracket}
      \end{cases}                   
\end{equation}
\noindent We also have the following correspondence for the physical quantities $f(\nu,x,y,z)$:
\begin{equation}  \label{eq:f_comput_domain}
      f_{\nu,z}\left(x_{i},y_{j}\right)_{\substack{\,i\in \llbracket N_{x}+1,N_{x}+i_{0} \rrbracket\\
                                                     j\in \llbracket N_{y}+1,N_{y}+j_{0} \rrbracket}}=
      f_{\nu,z}\left(x_{i'},y_{j'}\right)_{\substack{\,i'\in \llbracket 1,i_{0} \rrbracket\\
                                                       j'\in \llbracket 1,j_{0} \rrbracket}}
\end{equation}
\noindent We employ the long-characteristics method along $x$-direction from $(x_{i_{0}},y_{j_{0}})$ to $(x_{i_{0}+N_{x}},y_{j_{0}})$,
and along $y$-direction from $(x_{i_{0}},y_{j_{0}})$ to $(x_{i_{0}},y_{j_{0}+N_{y}})$.
Then, we resume our usual short-characteristics method for the next rows of the current $z$-layer.
Once calculated, the specific intensity is defined in the original computational domain, with a simple rearrangement of indices as per
Equation~(\ref{eq:f_comput_domain}).
Note that some long-characteristics may consist merely of short-characteristics. This is the case for a given grid point
$(x_{i,\,i\in \llbracket i_{0},N_{x} \rrbracket},y_{j_{0}})$, if the corresponding $\Delta x_{i}$ and $\Delta y_{\rm{max}}$ are large enough, so that
the related upwind end~point intersects the horizontal upwind face. Similarly, this is the case for
$(x_{i_{0}},y_{j,\,j\in \llbracket j_{0},N_{y} \rrbracket})$, if $\Delta y_{j}$ and $\Delta x_{\rm{max}}$ are large enough.

We mention here that it is also possible to handle periodic conditions by iterative methods, as done 
for instance by \cite{van_Noort_et_al_2002} and
\cite{Davis_et_al_2012}. However, such an approach can lead to a slow convergence in the case of optically thin media, as noted by
\cite{van_Noort_et_al_2002}.

\section{Tests of IRIS}  \label{sec:tests}

The code underwent a number of various tests to validate each of its functionalities that we describe in this paper.
We focus here on the three most important ones: the searchlight beam test, a comparison with a well-tested one-dimensional scheme,
and the test of the velocity gradient effect applied on one given spectral line.

\subsection{The Searchlight Beam Test}   \label{subsec:search_light_beam}
In the context of astrophysical radiation transport, this test was first proposed by \citet{Kunasz_and_Auer_1988}.
Several authors then used it to evaluate their multidimensional short-characteristics
algorithms \citep{Stone_et_al_1992c,Auer_and_Paletou_1994,Fabiani_Bendicho_2003,Fabiani_Bendicho_and_Trujillo_Bueno_2007,Hayek_et_al_2010,
Davis_et_al_2012}. The purpose is to examine how a beam propagates through a transparent medium.
Theoretically, the beam crosses the medium without being absorbed or dispersed. The RTE, Equation~(\ref{eq:RTE3}), is greatly simplified since, in
this particular case, $I_{c}=I_{u}$. Note that the numerical counterpart of the RTE, Equation~(\ref{eq:RTE4}) is consistent with this behavior when
the opacity is zero, as pointed out in Section~\ref{subsec:sc_discretized}.  Therefore, the calculation uses only the upwind interpolation of the
specific intensity. This test challenges the capabilities of the piecewise cubic, locally monotonic, interpolation technique
(see Section~\ref{sec:pcw_cubic_mon_intp}), as applied to cell face interpolations (see Section~\ref{subsec:intp_cell_faces}).

\begin{figure*} 
\centerline{
\includegraphics[width=1.00\columnwidth,angle=0,clip=true]{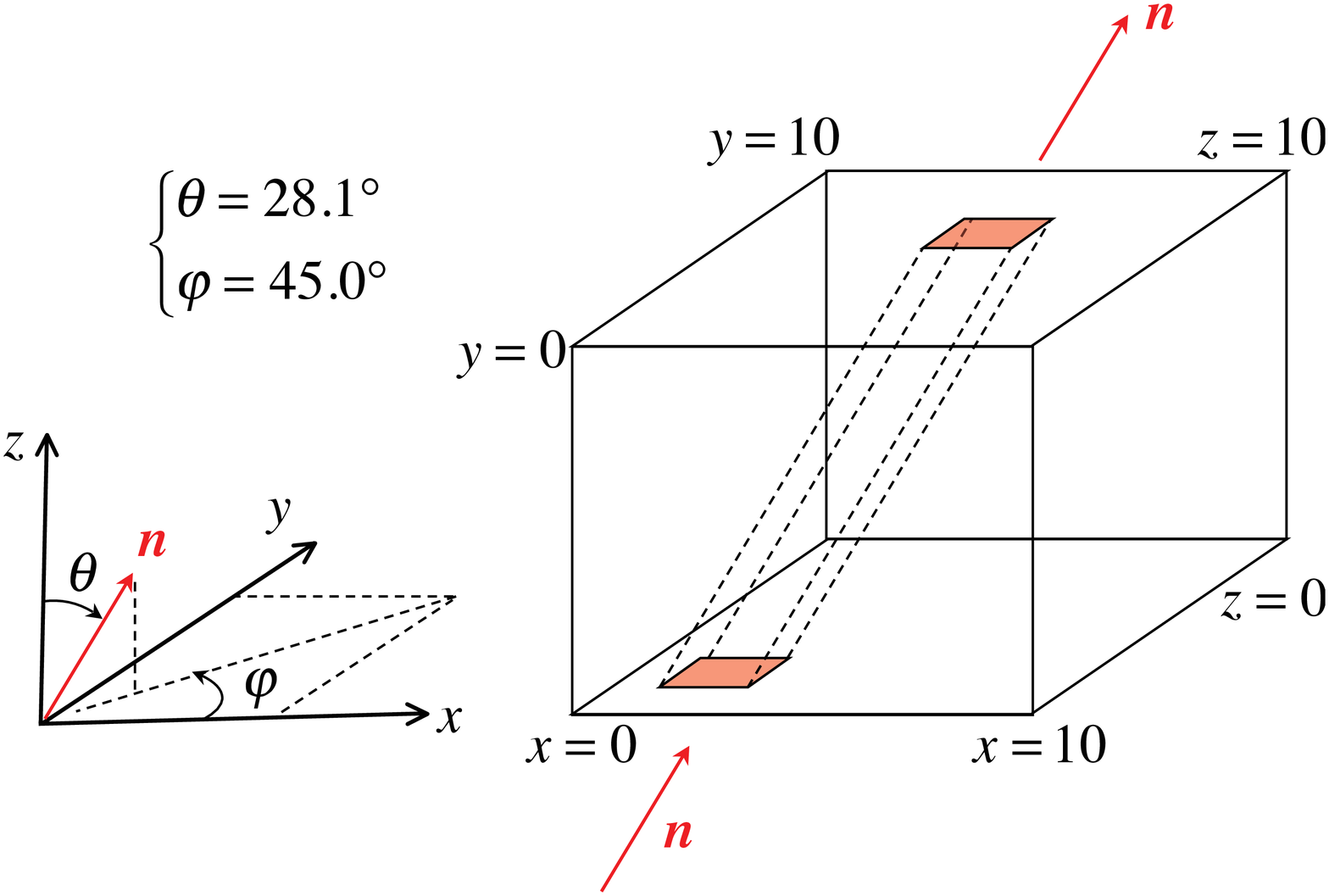}
\includegraphics[width=1.00\columnwidth,angle=0,clip=true]{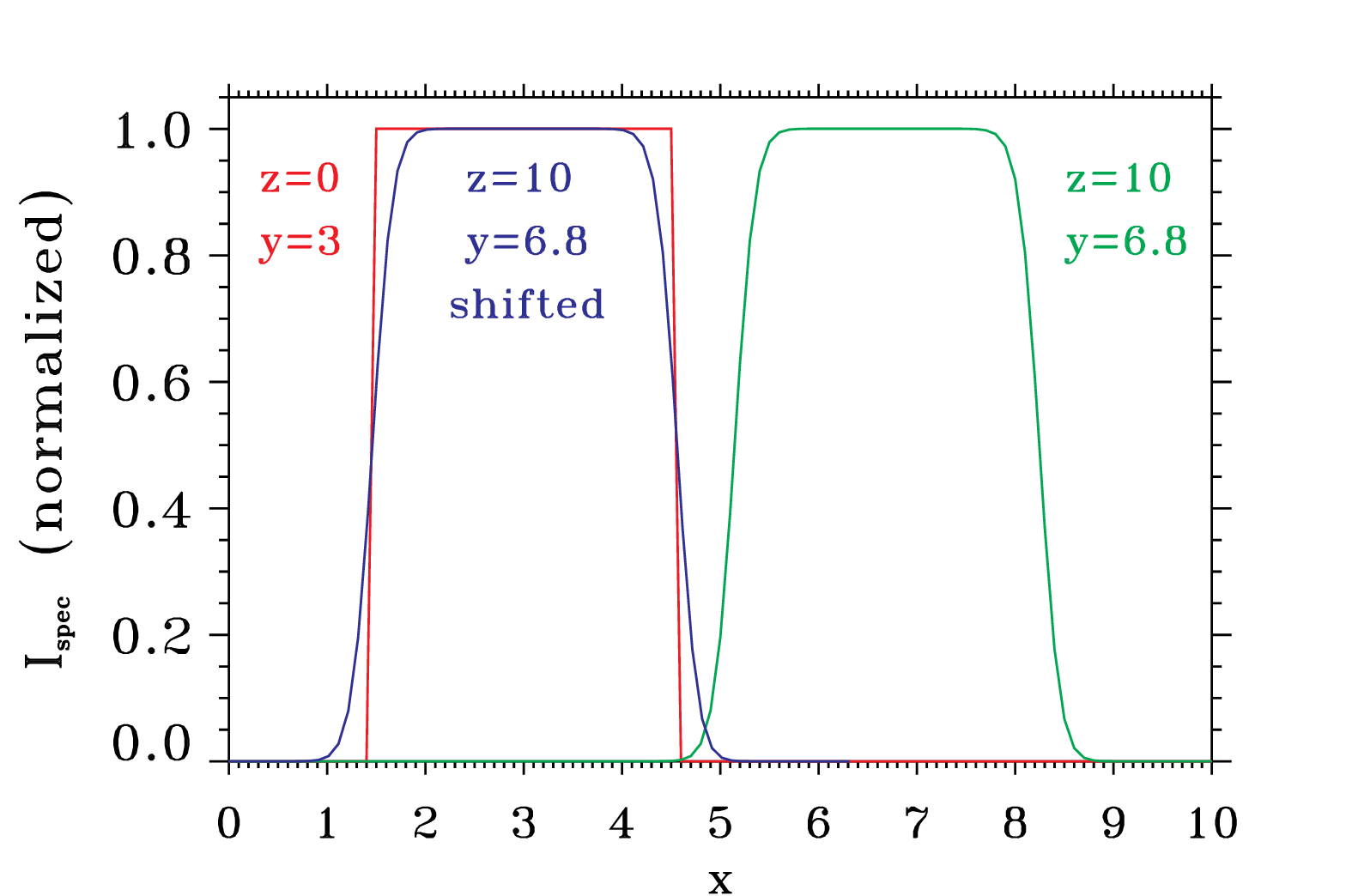}}
\centerline{
\includegraphics[width=1.00\columnwidth,angle=0,clip=true]{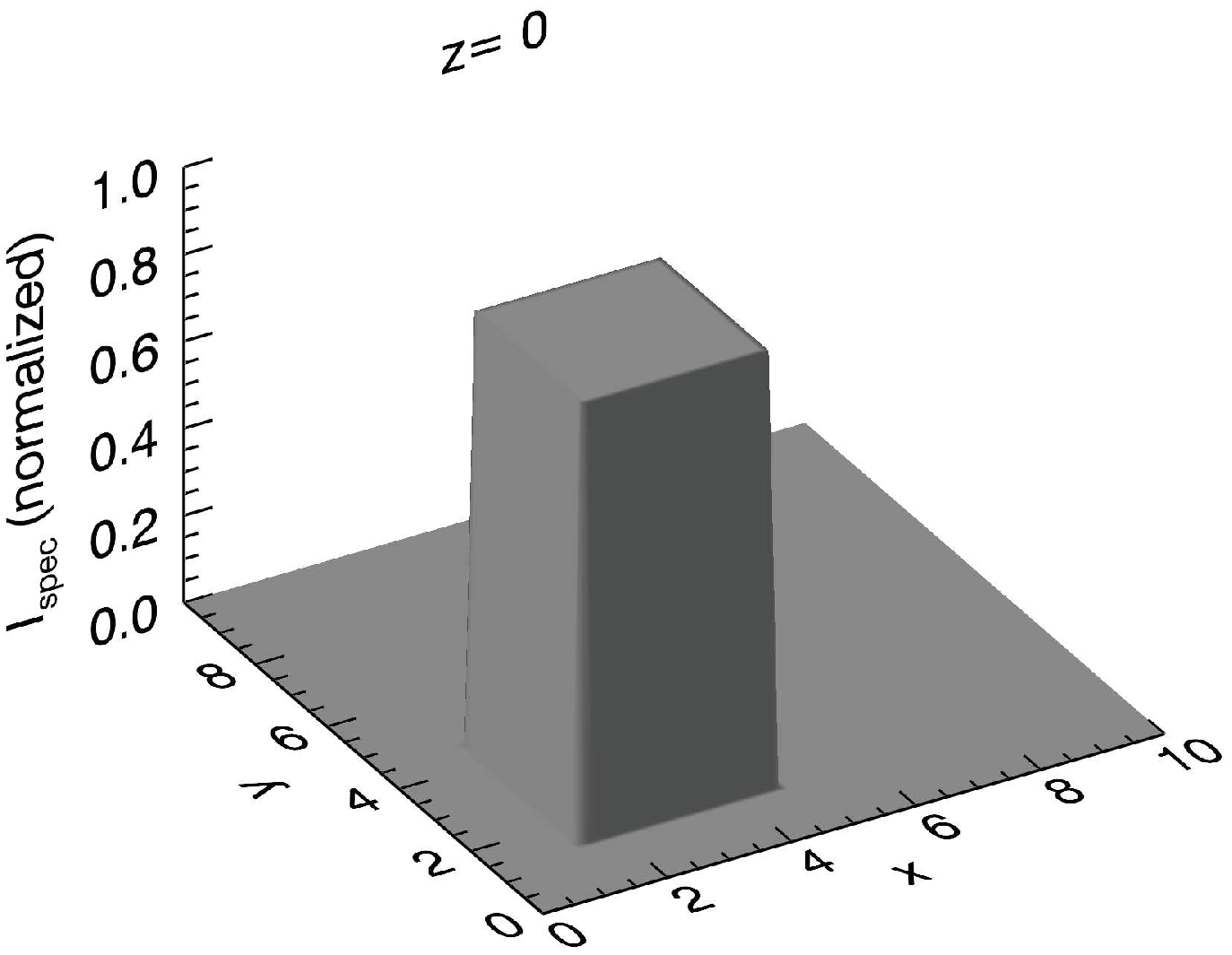}
\includegraphics[width=1.00\columnwidth,angle=0,clip=true]{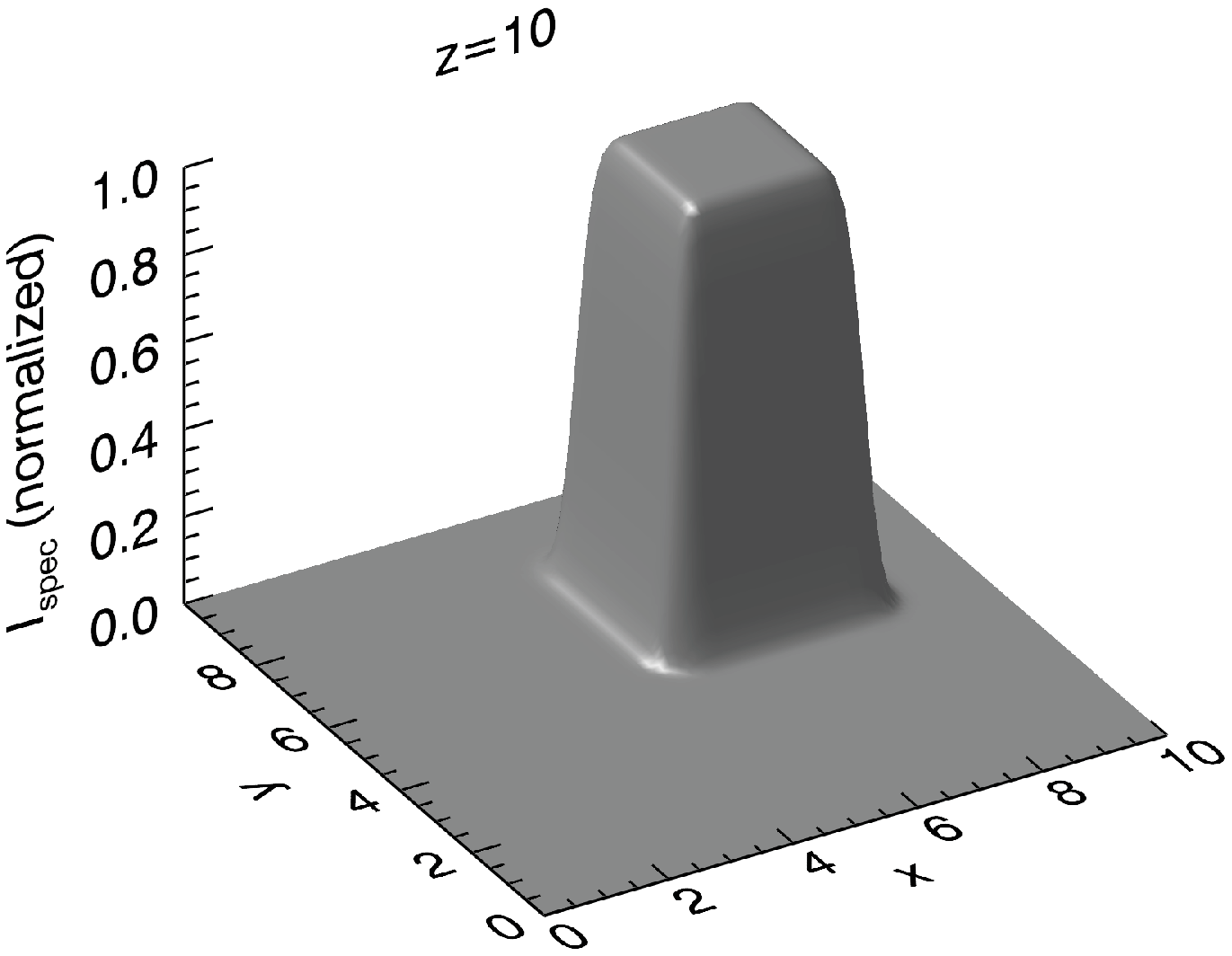}}
\caption{Searchlight beam test. The top left panel of the figure depicts the purpose of this test. A slanted beam, with direction $\bd{n}$, enters a
zero opacity box at its base, and emerges from it at the top. The objective is to compare the entering beam profile with the emerging one; both are
theoretically the same. The bottom panels show the normalized specific intensity of the beam, as a function of
positions $x$ and $y$, at the bottom of the box (left panel), and at the top of the box (right panel).
The top right panel displays sectional views of the beam along $x$-axis, at $z=0$ and $y=3$ (red curve), at $z=10$ and $y=6.8$ (green curve).
The blue curve is identical to the the green one, but shifted so that its center fits the center of the red curve.
See Section~\ref{subsec:search_light_beam} for explanations.}
\label{fig:SBT}
\end{figure*}

For comparison, the numerical test was performed with the same parameters as the ones used by \citet{Hayek_et_al_2010}. We consider a hard-edged beam
that propagates throughout a three-dimensional zero opacity box along a slanted direction defined by $\theta=28.1\degr$ and $\varphi=45.0\degr$, as
shown in the sketch (top left panel) of Figure~\ref{fig:SBT}. The box is made with $100^3$ grid points. It extends in each of the three directions
$(x,y,z)$ from 0 to 10 in arbitrary units. The beam is made with $30^2$ grid points at the base of the box. It enters the box at the base and emerges
from it at the top. The bottom panels of Fig.~\ref{fig:SBT} show the normalized specific intensity of the beam as a function of $x$ and $y$, at the
base of the box (left panel), and at the top of the box (right panel). We verify that the beam profile is very well conserved. There is no absorption:
the maximum value of the normalized specific intensity remains equal to one. The dispersion is very limited: the hard-edges are conserved, along with
the size of the beam. In addition, the beam exits at the right position of the box. The positions of the four corners of the beam,
$[(x_{1},y_{1}); (x_{2},y_{1}); (x_{1},y_{2}); (x_{2},y_{2})]$, are at the base:
$[(1.5,1.5); (4.5,1.5); (1.5,4.5); (4.5,4.5)]$, and at the top:
$[(5.3,5.3); (8.3,5.3); (5.3,8.3); (8.3,8.3)]$, which is consistent with
the $(\theta,\varphi)$ values specified above and the size of the box along $z$ direction.
In order to show in greater detail the dispersion of the beam, we have plotted sectional views of the beam along $x$-axis (top right panel).
The red curve represents the normalized specific intensity as a function of $x$, at $z=0$ (base of the box) and $y=3$ (middle of the beam).
The green curve represents this quantity at $z=10$ (top of the box) and $y=6.8$ (middle of the beam).
In order to compare the two profiles, we have plotted the blue curve, which is the green one artificially shifted so that its center fits the center
of the red curve. Such a superposition shows the symmetry of the beam's dispersion, which also remains small.
We verified that our scheme guarantees the photon conservation: the $xy$-surface integral of the specific intensity at $z=0$ equals the 
the $xy$-surface integral at $z=10$, to the machine accuracy.

This test represents an excellent validation of the piecewise cubic, locally monotonic, interpolation scheme that we use. Such a technique almost
suppresses the numerical diffusion effect of the short-characteristics method, and is more efficient than linear, parabolic (the first one is
extremely diffusive, the second one introduces spurious extrema, as shown by \citealt{Kunasz_and_Auer_1988}), or even monotonic
quadratic schemes \citep{Auer_and_Paletou_1994}.
 
\subsection{Comparison with 1D Plane-Parallel Models}  \label{subsec:cmp_3D_1D}
It is important to check whether IRIS can reproduce the results provided by one-dimensional radiative transfer schemes, when the medium is made of
homogeneous plane-parallel layers, i.e., in the case of a 1D plane-parallel structure.
Simulating the radiative transfer for such a medium with IRIS can be achieved by imposing periodic boundary conditions in faces perpendicular to
the parallel layers. Assuming that these layers are parallel to the horizontal $z$-planes, the specific intensity has then the following
dependence, $I(z,\cos \theta,\nu)$. Applying the full 3D calculation with IRIS for such a medium perfectly reproduces the radiative results provided
by 1D radiative models at any frequency, as we show in the example below.

We calculate the radiative transfer through a radiative shock structure made of homogeneous plane-parallel layers. The hydrodynamics structure,
provided by Matthias Gonz{\'a}lez\footnote{AIM, CEA/DSM/IRFU, CNRS, Universit{\'e} Paris Diderot, 91191 Gif-sur-Yvette, France}, is obtained with the
three-dimensional radiation hydrodynamics code HERACLES \citep{Gonzalez_et_al_2007}. The test case is a simulation of an experimental radiative shock
generated in a tube full of Xenon assumed to be a perfect gas, with the following upstream conditions: fluid velocity = 60~$\rm km~s^{-1}$,
pressure~=~7~bar, temperature = 1~eV.  The opacities that we use are from \citet{Michaut_et_al_2004}. Our objective here is focused on the comparison
of the 1D and the 3D radiative models; a future paper (Ibgui et al.~2012, in preparation) will present
more details of three-dimensional models of radiative shocks.

\begin{figure*} 
\centerline{
\includegraphics[width=1.0\columnwidth,angle=0,clip=true]{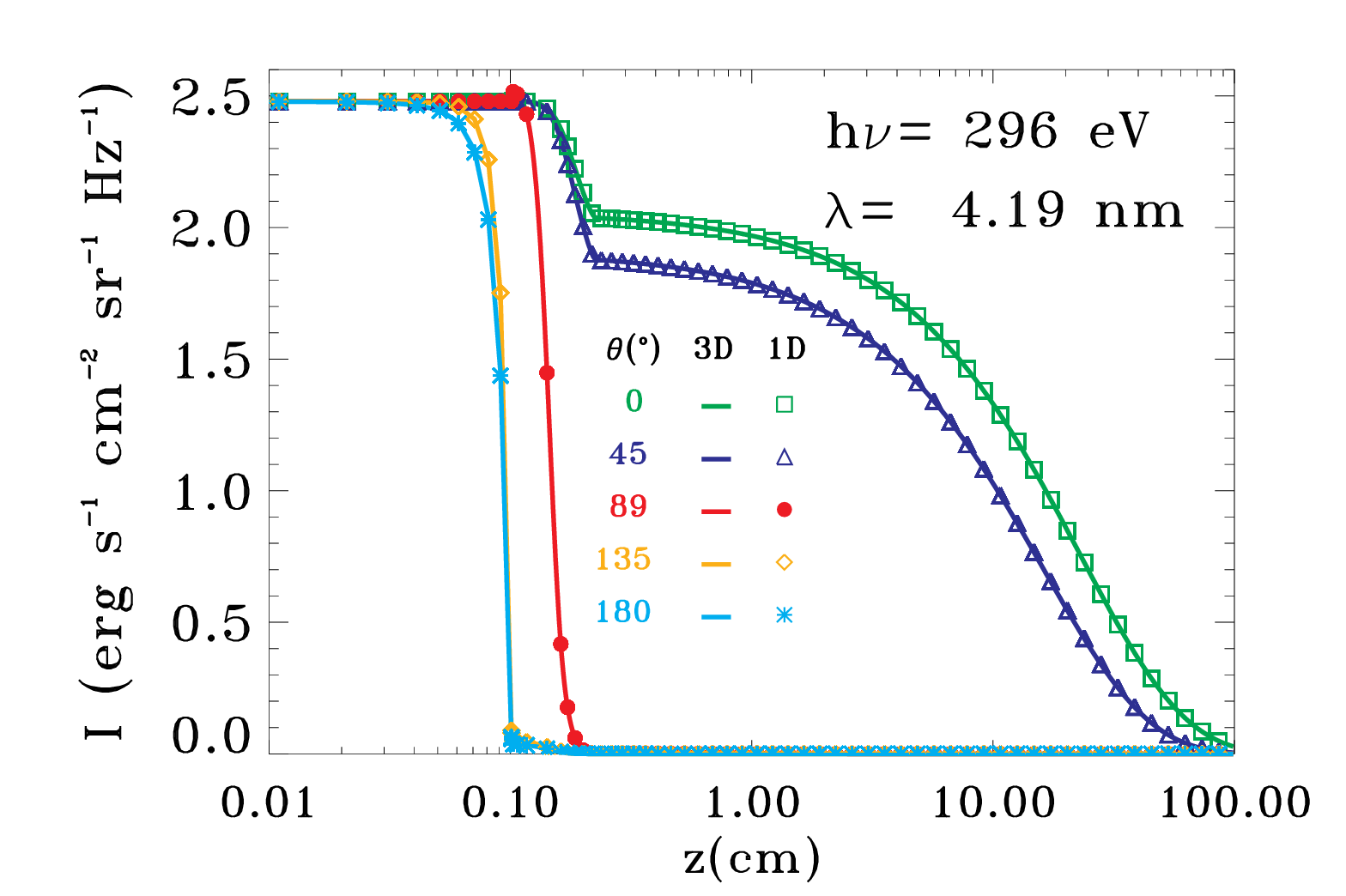}
\includegraphics[width=1.0\columnwidth,angle=0,clip=true]{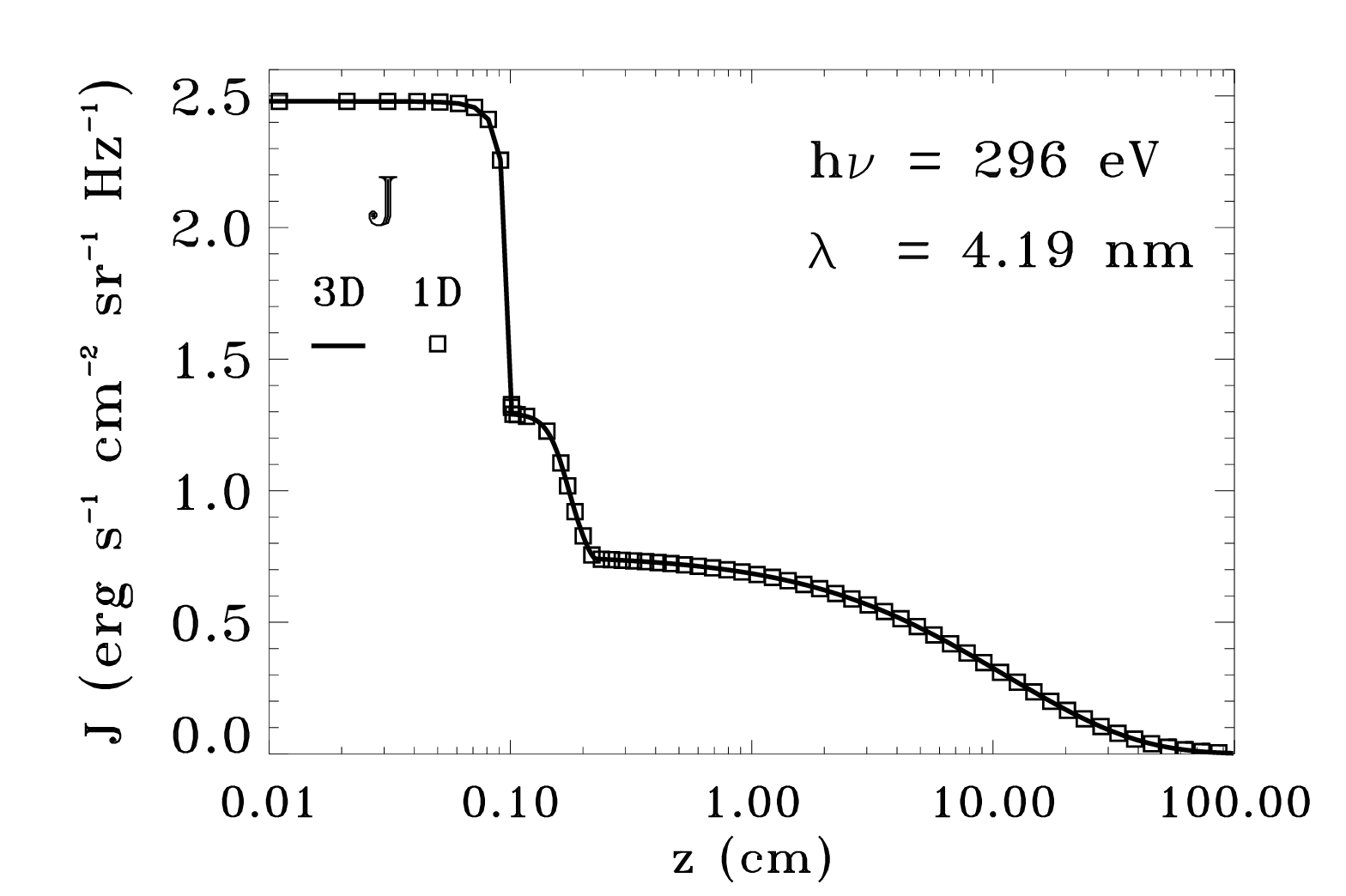}}
\centerline{
\includegraphics[width=1.0\columnwidth,angle=0,clip=true]{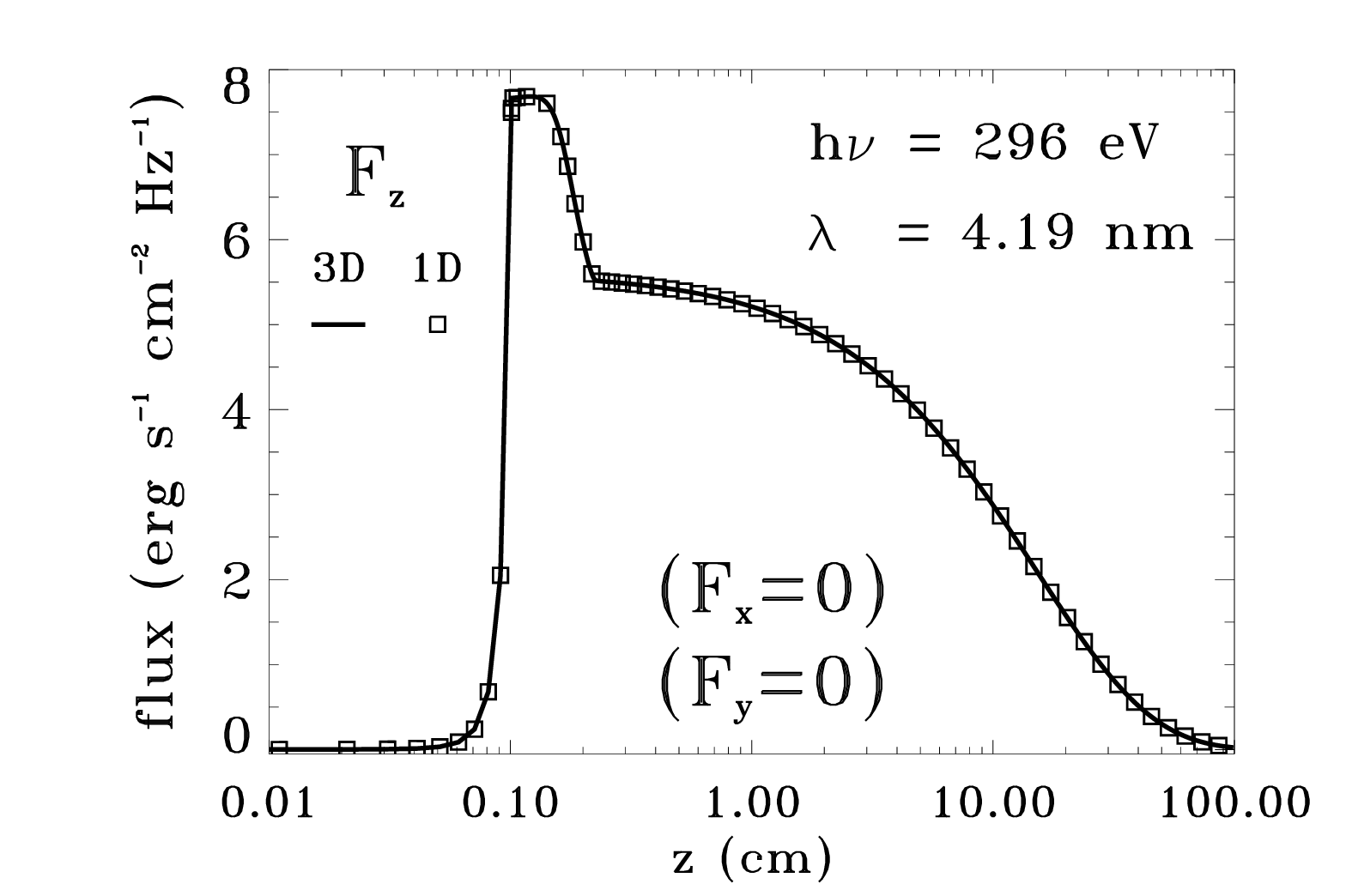}
\includegraphics[width=1.0\columnwidth,angle=0,clip=true]{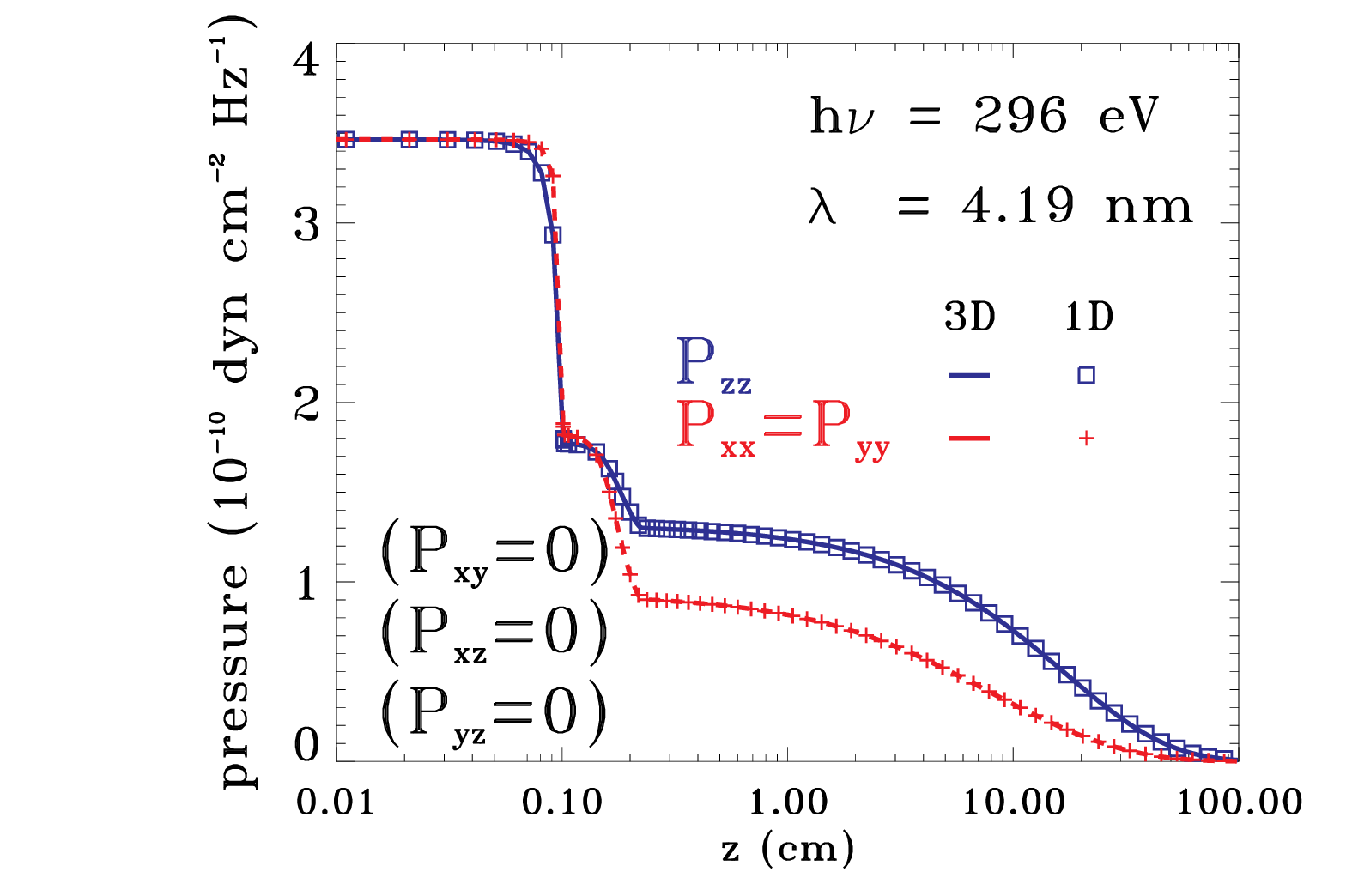}}
\caption{Comparison with 1D plane-parallel models. The figures shows the monochromatic radiative quantities, at $h\nu = 296\rm~eV$ 
($\lambda=4.19~\rm nm$), calculated with IRIS (solid lines) and with a 1D solver (symbols), in the case of a radiative shock structure made of homogeneous plane-parallel
layers. The top left panel displays the specific intensity for different polar angles $\theta$: $0\degr$~(green, square), $45\degr$~(blue, triangle), $89\degr$~(red, filled circle),
$135\degr$~(orange, diamond), $180\degr$~(cyan, asterisk), and for a set of twelve azimuthal angles
$\varphi(\degr)=(0, 45, 89, 90, 135, 179, 180, 225, 269, 270, 315, 359)$. There is no legend for the
$\varphi$ values, because all the curves, for a given $\theta$, are perfectly superposed. The top right, bottom left, and bottom right panels show
respectively the mean intensity $J$~(black, square), the components of the radiative flux vector $\bd F$~(black, square), and the components of the radiation pressure tensor
$\bd{\mathsf P}$ ($P_{zz}\text{: blue, square; } P_{xx}=P_{yy}\text{: red, plus}$). See Section~\ref{subsec:cmp_3D_1D} for more details.}
\label{fig:cmp_3D_1D}
\end{figure*}

The 3D computational grid is built with $20\times20\times510$ points in respectively $x$, $y$, and $z$ directions. In each horizontal $z$-plane, all
the state parameters (temperature, density, velocity) are independent of $x$ and $y$. Two post-processing calculations were performed: one
with IRIS applied on the 3D grid, and one with a well-tested 1D short-characteristics solver applied on a 1D grid that is extracted from the 3D one.
Both were done at seventeen frequencies that range from $h\nu = 2\rm~eV$ ($\lambda=620~\rm nm$) to $h\nu = 494\rm~eV$ ($\lambda=2.41~\rm nm$).
The velocity gradient effect was not taken into account in this test.
Figure~\ref{fig:cmp_3D_1D} shows several radiative quantities provided by IRIS and by the 1D solver.
By way of example, we select one frequency, $h\nu = 296\rm~eV$, which corresponds to $\lambda=4.19~\rm nm$.

The top left panel of Figure~\ref{fig:cmp_3D_1D} displays the specific intensity as a function of $z$, and for different polar angles $\theta$:
$0\degr$~(green), $45\degr$~(blue), $89\degr$~(red), $135\degr$~(orange), $180\degr$~(cyan). The solid lines are the results calculated with IRIS.
The symbols are the results provided by the 1D~solver. Note that, for the readability of the figure, we plot the 1D results for only some of the
510 $z$ values. Note also that we plot the 3D results calculated with several azimuthal angles
$\varphi(\degr)=(0, 45, 89, 90, 135, 179, 180, 225, 269, 270, 315, 359)$. All the curves for these twelve $\varphi$ angles are perfectly superposed,
and, therefore, indistinguishable. This demonstrates that the specific intensity is independent of $\varphi$. In addition, we verified that for any
given pair $(\theta,\varphi)$, the specific intensity at a given $z$-plane is the same for all the grid points of this plane, which demonstrates that
this quantity is independent of $x$ and $y$. Last but not least, the 1D symbols perfectly fit with the 3D curves.

The other three panels of Figure~\ref{fig:cmp_3D_1D} display the moments obtained by angular integration (see
Section~\ref{sec:moments_angular_integration}): the mean intensity $J$ in the top right panel, the components of the radiation flux vector, $F_{x}$,
$F_{y}$, $F_{z}$, in the bottom left panel, and the components of the radiation pressure tensor $P_{xx}$, $P_{yy}$, $P_{zz}$, $P_{xy}$, $P_{xz}$,
$P_{yz}$, in the bottom right panel. All the curves show that the 1D and 3D results perfectly agree.
We checked this result for all the
seventeen tested frequencies. In addition, the 3D results provided by IRIS verify, for any frequency, the following properties that are
theoretically valid for 1D plane-parallel structures. The radiative flux is zero in $x$ and $y$ direction, $F_{x}=0$, $F_{y}=0$, and it depends only
on $z$ in $z$-direction, $F_{z}(z,\nu)$. The non-diagonal components of the radiation pressure tensor are all zero: $P_{xy}=0$, $P_{xz}=0$,
$P_{yz}=0$. The other three pressure components depend only on $z$, $P_{xx}(z,\nu)$, $P_{yy}(z,\nu)$, $P_{zz}(z,\nu)$, and we have, for any $z$ and
$\nu$, $P_{xx}(z,\nu)=P_{yy}(z,\nu)$.

These tests demonstrate that IRIS is capable of reproducing 1D plane-parallel simulations, and, thereby, of handling periodic boundary conditions. 

\subsection{Test of the Velocity Gradient Effect}  \label{subsec:velocity_effect_1_line}
As explained in Section~\ref{subsec:velocity_effect}, IRIS can handle the velocity gradient effect on spectra (caused by the Doppler shifts of photon
frequencies) by subdividing a short-characteristic in a set of subintervals. This subdivision is controlled by a tunable parameter $\epsilon_{\sm D}$,
so that the Doppler shift of any spectral line between two subinterval points remains bounded by a fraction $\epsilon_{\sm D}$ of the local Doppler
width of the line, as per Equation~(\ref{eq:delta_nu_shift}).

\begin{figure*} 
\centerline{
\includegraphics[width=1.0\columnwidth,angle=0,clip=true]{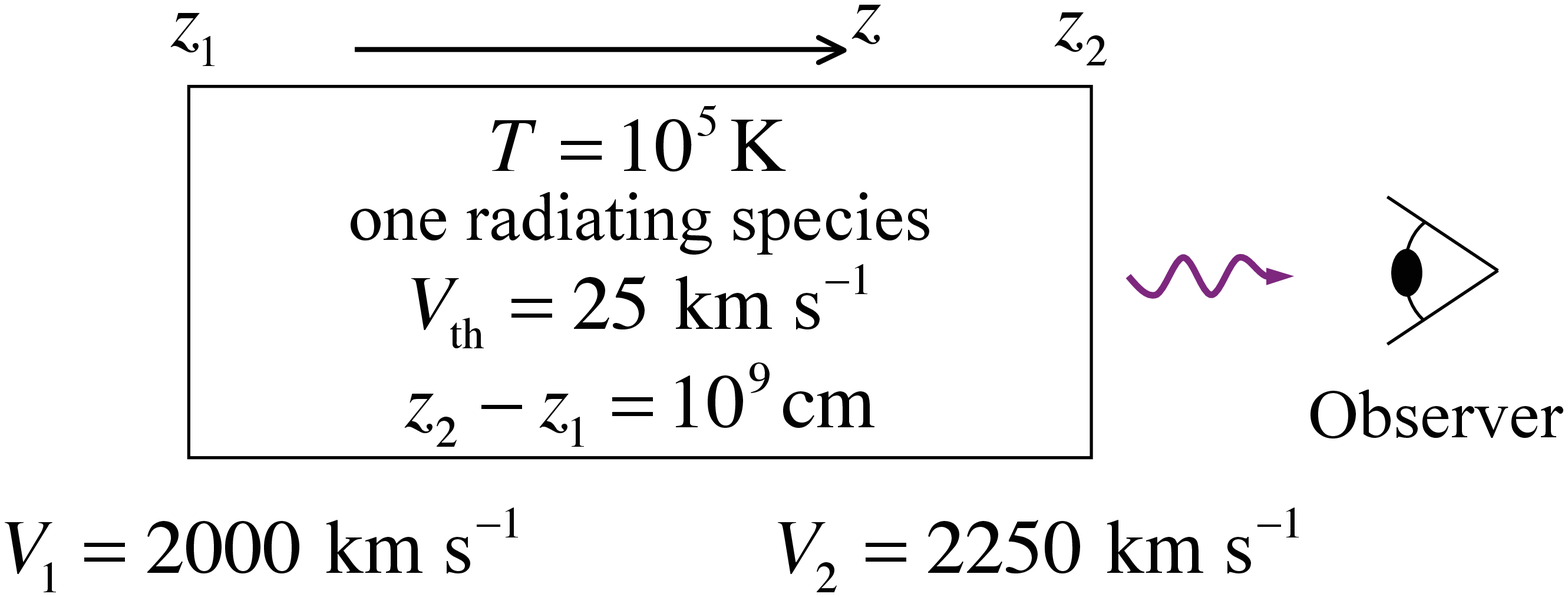}}
\centerline{
\includegraphics[width=1.0\columnwidth,angle=0,clip=true]{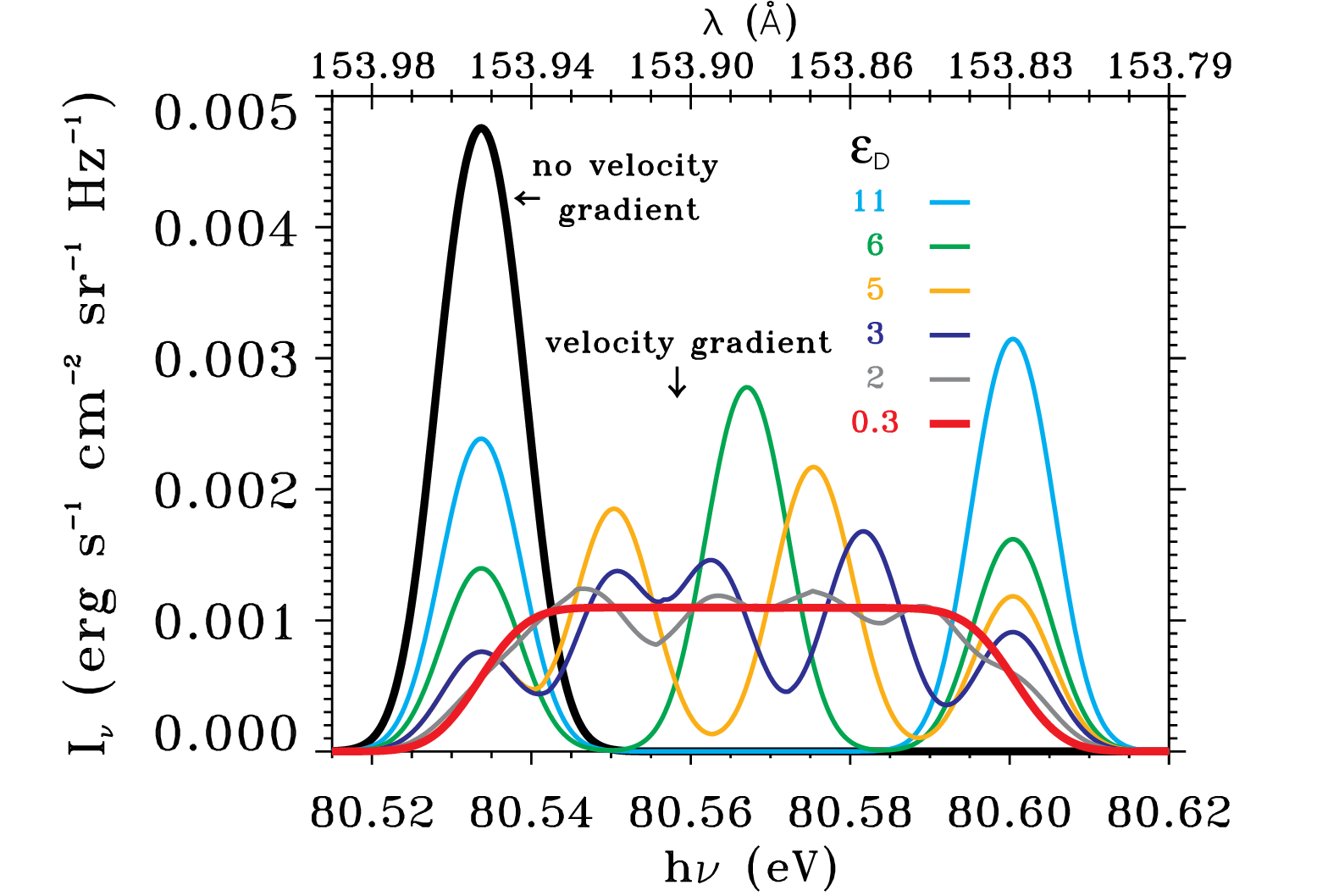}
\includegraphics[width=1.0\columnwidth,angle=0,clip=true]{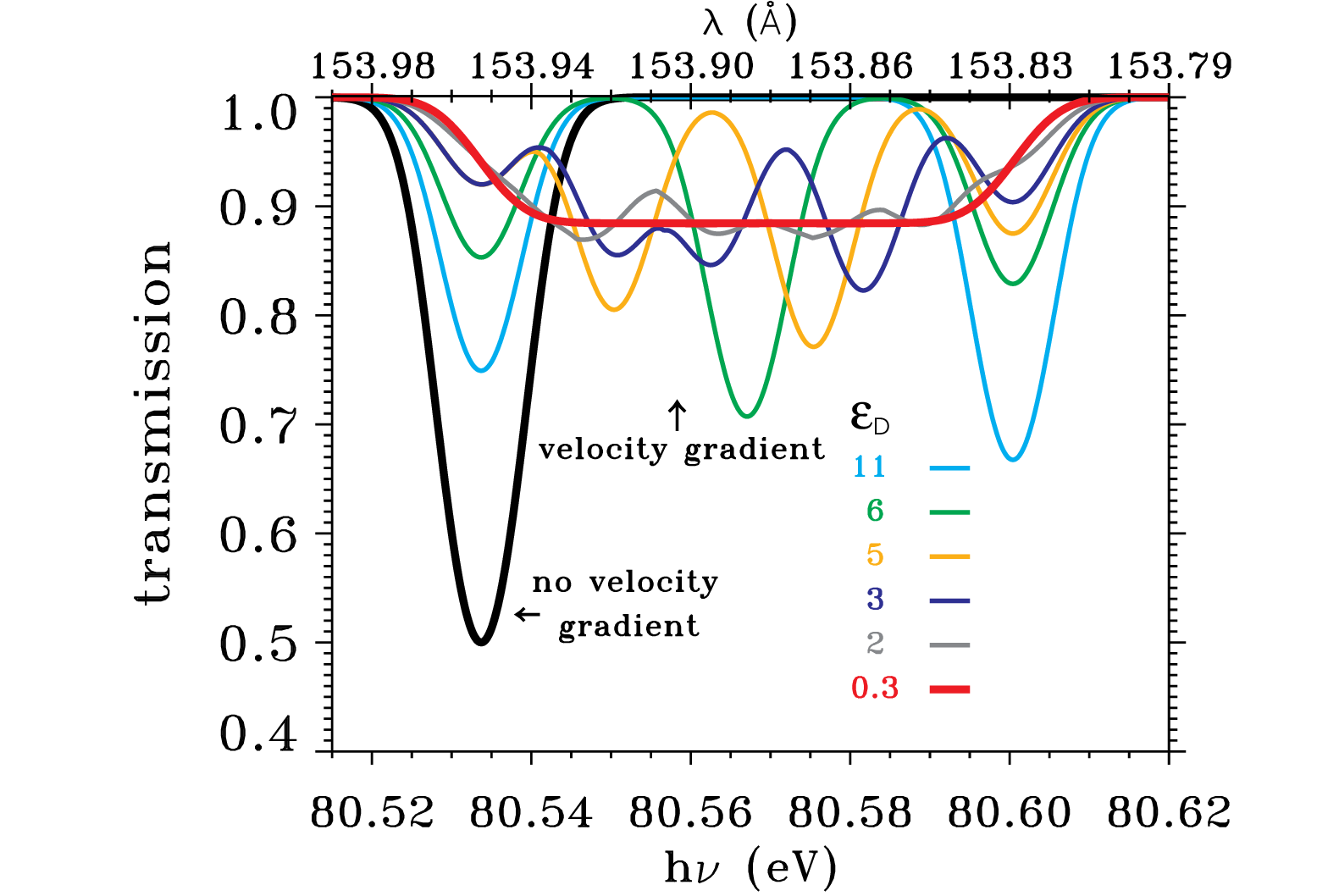}}
\caption{Test of the velocity gradient effect. The top panel represents a single plasma layer in a stellar wind, with a velocity gradient. The layer
is at uniform temperature $T$, with one radiating species, assuming LTE and no scattering. The layer is modeled in IRIS with one cell between grid
points $z_{1}$ and $z_{2}$, with wind velocities $V_{1}>0$ (the material approaches the observer) and $V_{2}>V_{1}$.
The red thick curve in the bottom left panel displays the monochromatic specific intensity $I_{\nu}$, due to the radiation that emerges from the layer
at $z_{2}$ toward the observer, as computed by IRIS. This emission spectrum is plotted versus photon energy $h\nu$~(eV) or photon wavelength
$\lambda~(\AA)$. It is obtained with a parameter $\epsilon_{\sm D}=0.3$. Too large values of $\epsilon_{\sm D}$ result in incorrect profiles, as shown
by the following curves: cyan~($\epsilon_{\sm D}=11$), green~($\epsilon_{\sm D}=6$), orange~($\epsilon_{\sm D}=5$), blue~($\epsilon_{\sm D}=3$),
grey~($\epsilon_{\sm D}=2$).
The bottom right panel displays the corresponding transmission spectrum.
For comparison, the black thick curve portrays the spectral line that would obtain if there was no velocity gradient ($V_{2}=V_{1}$).
See Section~\ref{subsec:velocity_effect_1_line} for detailed explanations.}
\label{fig:velocity_effect_1_line}
\end{figure*}

We illustrate the effect of our treatment in a simple ideal case that can be otherwise calculated quasi-analytically.
We consider one plasma layer with a uniform temperature $T$ in a stellar wind region with a velocity gradient, as sketched by
Figure~\ref{fig:velocity_effect_1_line}, top panel. We focus on one velocity direction, referred to as $z$-direction. In our example, the material
moves toward an observer.  
The layer is numerically modeled with a single cell between two grid points, $z_{1}$, the farther position from the observer, and $z_{2}$, the closer
position to the observer. The wind velocities at grid points are $V_{1}>0$ and $V_{2}>V_{1}$. 
We assume that there is only one radiating species. We focus on the radiation of the layer itself, in the direction of the wind toward the observer,
and neglect any incoming radiation in $z_{1}$. 
Although the situation in a stellar wind is significantly different from LTE, we do assume LTE for the purpose of this test. We also neglect
scattering. Therefore, the source function is uniform and equals the Planck function at the layer's temperature, $B_{\nu}\left(T\right)$.
We consider one spectral line, for which we assume, for simplicity, a Doppler profile centered at a frequency $\nu_{0}$ in the particle's frame. With
these assumptions, the Doppler width $\Delta \nu_{\sm D}$ is a constant, and the monochromatic specific intensity of the radiation emerging
from the layer toward the observer is simply given by:
\begin{equation}  \label{eq:RTE_velocity_effect}
   I(\nu) = \left(1-e^{-\tau_{12}(\nu)}\right) \ B_{\nu}(T) \:
\end{equation}
where $\tau_{12} (\nu)$ is the monochromatic optical depth from $z_{1}$ to $z_{2}$:
\begin{equation}  \label{eq:tau12_velocity_effect}
   \tau_{12} (\nu) = \int_{z_{1}}^{z_{2}}\kappa(\nu,z)\ dz \:,
\end{equation}
where the absorption coefficient $\kappa(\nu,z)$ is written as 
\begin{equation}  \label{eq:kappa_velocity_effect}
   \kappa(\nu,z) = K \ e^{- \left( \frac{\nu-\nu_{\rm ctr}(z)}{\Delta \nu_{\sm D}}\right)^{2} } \:,
\end{equation}
where $\nu_{\rm ctr}(z)$, defined in Section~\ref{subsec:velocity_effect} by Equation~(\ref{eq:nu_center}), is the shifted line-center frequency for a
velocity $V(z)$ at position $z$:  $\nu_{\rm ctr}(z)~=~\nu_{0}~\left( 1+ V(z)/c \right)$.
The quantity $K$ is a constant that contains all the attributes of the line (population of the lower level, and
oscillator strength) and the value of the constant Doppler width.

Now, since we use monotonic laws (cubic Hermite polynomials, see Section~\ref{subsec:sc_discretized}) to interpolate physical quantities along a 
short-characteristic, $\tau_{12} (\nu)$ may be written as 
\begin{equation}  \label{eq:tau12_convolution_1}
  \tau_{12} (\nu) = \int_{\nu_{\rm ctr_{1}}}^{\nu_{\rm ctr_{2}}} R'(\nu_{\rm ctr}) \ K \ e^{- \left( \frac{\nu-\nu_{\rm ctr}}{\Delta \nu_{\sm D}}\right)^{2} }\ d\nu_{\rm ctr} \:,
\end{equation}
where $R$ is the function relating the shifted transition frequency $\nu_{\rm ctr}$ to the position $z$ within the layer: $z=R(\nu_{\rm ctr})$,
$R'$ being its derivative, and where we define:
\begin{equation}  \label{eq:nu_ctr_1_2}
   \begin{cases}
      \nu_{\rm ctr_{1}}=\nu_{0} \left( 1+ \tfrac{V_{1}}{c} \right) \\
      \nu_{\rm ctr_{2}}=\nu_{0} \left( 1+ \tfrac{V_{2}}{c} \right)
   \end{cases}
\end{equation}

In this simplified configuration, the optical depth appears as the
convolution of a function $A(\nu)$ and a Gaussian function $B(\nu)$ (the line profile), both defined as follows:
\begin{subequations}  \label{eq:A_B_convolution}
      \begin{equation}  \label{eq:tau12_convolution_2}
         \tau_{12} (\nu) = (A \ast B)\ (\nu) \:,
      \end{equation}
      \text{where} 
      \begin{equation}  \label{eq:A}
         A(\nu) =
         \left\{
         \begin{alignedat}{3}
             &R'(\nu_{\rm ctr}) & \quad \text{if} & \quad \text{$\nu_{\rm ctr_{1}} \leqslant \nu \leqslant \nu_{\rm ctr_{2}}$} & \\
             &0                 & \quad \text{if} & \quad \text{$\nu < \nu_{\rm ctr_{1}}$  or \ $\nu > \nu_{\rm ctr_{2}}$}
         \end{alignedat}
         \right.
      \end{equation}
      \begin{equation}  \label{eq:B}
         B(\nu) = K e^{-\nu^2}
      \end{equation}
\end{subequations}

For the test, we adopt the following typical values of velocities in winds of early massive stars: $V_{1}=2000~\rm{km~s^{-1}}$,
$V_{2}=2250~\rm{km~s^{-1}}$, and a thermal velocity (constant in our uniform layer) of $V_{\rm th}=25~\rm{km~s^{-1}}$.
With such a choice, any line is Doppler shifted by an amount of 10~times its Doppler width between position $z_{1}$ and
position $z_{2}$, i.e., $(\nu_{\rm ctr_{2}}-\nu_{\rm ctr_{1}})/\Delta \nu_{\sm D}=(V_{2}-V_{1})/V_{\rm th}=10$.
We also adopt a temperature $T=10^{5}$~K, and a layer size $z_{2}-z_{1}=10^{9}$~cm.

We consider an artificial line, arbitrarily centered at $h\nu_{0}=80.00$~eV in the particle's frame. In the observer's frame, the line is centered at
$\nu_{\rm ctr_{1}}\approx 80.53$~eV at $z_{1}$, and at $\nu_{\rm ctr_{2}}\approx 80.60$~eV at $z_{2}$.
Note also that we have defined the $K$ value (see Equation~\ref{eq:kappa_velocity_effect}) so that the transmission of the layer at the line center
equals $0.5$ in the case when there is no velocity gradient. In this case indeed, the monochromatic thermal absorption $\kappa(\nu,z)$ is
independent of $z$, so that $K$ is inferred from $e^{-K(z_{2}-z_{1})}=0.5$.
This way, we avoid the extreme cases of an optically thin or an optically thick layer.
The red thick curve in Figure~\ref{fig:velocity_effect_1_line}, bottom left panel, displays the monochromatic specific intensity $I(\nu)$ that emerges
from the layer at position $z_{2}$ toward the observer, as calculated by IRIS. This emission spectrum is displayed versus photon energy $h\nu$~(eV)
or photon wavelength $\lambda~(\AA)$.
For comparison, the black thick curve represents the line, centered at 80.53~eV, that would obtain if there was no velocity gradient, i.e., if
$V_{2}=V_{1}$.
The bottom right panel of the figure portrays the transmission, $e^{-\tau_{12}(\nu)}$, of the layer.
In the highly simplified configuration of our test, the transmission and the specific intensity are related by the linear
Equation~(\ref{eq:RTE_velocity_effect}). This is why this transmission spectrum looks the same, but inverted, as
the emission spectrum. In the presence of a velocity gradient, the absorption of the medium decreases, is shifted (blueward if the material approaches
the observer, redward if it recedes from the observer), and spreads over a larger spectral range. This also applies to the emission of the medium.

In this ideal test case, we can explain the shape and the position of the line depicted by the red curve.
The optical depth is given by Equation~(\ref{eq:tau12_convolution_1}). In addition, the layer is modeled with a unique cell.
So, the cubic monotonic Hermite polynomial that defines $V(z)$ along the short-characteristic between $z_{1}$ and $z_{2}$ almost coincides with a
linear law, since the downwind endpoint is a ghost point in which state parameters are defined by linear extrapolations (see
Section~\ref{subsec:intp_cell_faces}).
In these specific conditions, $z$ is trivially related to $\nu_{\rm ctr}$ by a linear function $R(\nu_{\rm ctr})$,
so that $A(\nu)$ is merely a boxcar function.
Then, the optical depth is
\begin{equation}  \label{eq:tau12_convolution_linear}
  \tau_{12} (\nu) = K \frac{c}{\nu_{0}} \frac{z_{2}-z_{1}}{V_{2}-V{z_{1}}} \int_{\nu_{\rm ctr_{1}}}^{\nu_{\rm ctr_{2}}} e^{- \left( \frac{\nu_{\rm ctr}-\nu}{\Delta \nu_{\sm D}}\right)^{2} }\ d\nu_{\rm ctr}
\end{equation}
where $\nu_{\rm ctr_{1}}$ and $\nu_{\rm ctr_{2}}$ are defined by Equations~(\ref{eq:nu_ctr_1_2}).
The calculation of the monochromatic specific intensity with Equation~(\ref{eq:RTE_velocity_effect}), using the
relation~(\ref{eq:tau12_convolution_linear}) above, provides a result that exactly fits the red curve obtained with IRIS.
Note that this spectral line is centered at $(\nu_{\rm ctr_{1}}+\nu_{\rm ctr_{2}})/2 \approx 80.57$~eV.

Last but not least, Fig.~\ref{fig:velocity_effect_1_line} (emission spectrum, left, or transmission spectrum, right) shows that the
$\epsilon_{\sm D}$ parameter (cf. Equation~(\ref{eq:delta_nu_shift})) plays a crucial role in the precision of the calculated line, through the
subgriding of the short-characteristic.
The cyan curve shows the profile obtained with $\epsilon_{\sm D}=11$, for
which we checked that the short-characteristic is not subdivided. This profile is composed of two peaks, which are located in the two extreme
positions of the line center, $\nu_{\rm ctr_{1}}$ and $\nu_{\rm ctr_{2}}$, with a large trough in the middle. Such a discrepancy between this profile
and the correct red one may result in erroneous interpretations of observed spectra when compared with synthetic spectra.
As $\epsilon_{\sm D}$ decreases, the trough is being gradually filled, as shown by the other curves obtained with different $\epsilon_{\sm D}$:
6~(green), 5~(orange), 3~(blue), 2~(grey).
This test verifies that the line shape starts to stabilize from $\epsilon_{\sm D}=1$. Tiny waves remain in the profile calculated with
$\epsilon_{\sm D}=1$. This profile is not shown in the figure because these waves are barely perceptible, so that it almost coincides with
the red profile.
Our numerical tests indicate that the short-characteristic is subdivided in 14 subintervals for $\epsilon_{\sm D}=1$, whereas it is subdivided
in 46 subintervals for $\epsilon_{\sm D}=0.3$. Not surprisingly, increasing the precision involves a larger computational cost. Such a test confirms
that the ideal value of $\epsilon_{\sm D}$ lies somewhere between 1/3 and 1 (see Section~\ref{subsec:velocity_effect}).

We stress again that the profile depicted by the red curves in Fig.~\ref{fig:velocity_effect_1_line} has been obtained in the ideal case of a single
line with a Doppler profile, under the assumption of a unique layer at uniform temperature, in LTE, and with a linear velocity variation within the
layer. In the general case, none of these situations holds, which results in a more complex profile. Only a full calculation with our subgriding
method can provide the actual profile.

\section{Conclusion}  \label{sec:conclusion}

In this paper, we have described IRIS, a new generic three-dimensional spectral radiative transfer code, which solves the monochromatic static
3D radiative transfer equation in the observer's frame, in Cartesian coordinates. We drop the time derivative in the transfer equation because we
consider situations in which the dynamical timescales are large compared to the photon free-flight time, such that the studied medium is assumed to
be in a frozen state during the time when the radiation field propagates throughout its structure.
The code is primarily intended for post-processing any hydrodynamics snapshot, i.e., any structure provided at a given instant by a
(radiation)(magneto)hydrodynamics simulation. That way, IRIS can determine, at each instant, the radiation field of a non-stationary flow.
We consider non-relativistic flow velocities. Currently, we assume local thermodynamic equilibrium.

IRIS is mainly a diagnostic tool to be used for comparisons between predicted spectra, maps or images one the one hand, and astrophysical
observations or laboratory astrophysics measurements on the other hand. It computes the monochromatic specific intensity and optical depth at any
position within a hydrodynamics structure, for any direction.
Associated with appropriate opacities specified by the user in a
dedicated module, it calculates synthetic spectra and maps or images emerging from the studied structure or astrophysical object. It can also 
calculate, from the specific intensity and through angular integrations, the moments of the radiation field: the mean intensity,
the radiation flux vector, and the radiation pressure tensor, at any position within the studied medium.

IRIS works with any 3D Cartesian grid, the latter being nonuniform in each direction.
The code uses a short-characteristics solver to
determine the formal solution of the radiative transfer equation. We have implemented a very efficient piecewise cubic, locally monotonic,
interpolation technique, that considerably reduces the numerical diffusion effects of the short-characteristics method. The latter is used for
interpolating any physical quantity in the cell faces of the computational grid, and for defining laws
of variation of physical quantities along short-characteristics.

IRIS is able to handle horizontal periodic boundary conditions of a simulation box. This configuration occurs for a medium with an infinite
(that is, large compared to the extent of the computational box)
extension in its horizontal plane and a double periodicity in this plane, along with a finite extension in its vertical direction, such as a
stellar atmosphere or an accretion disk.

Since it is formulated in the observer's frame, IRIS can deal with any (non-relativistic) 
non-monotonic macroscopic velocities. 
The code can be applied to a large number of radiating astrophysical objects or structures, such as
accretion shocks or jets in young stellar objects, stellar atmospheres, exoplanet atmospheres, accretion disks, rotating stellar winds, and
cosmological structures. IRIS has already been applied to predict X-UV spectra of laboratory generated radiative shocks
(see conference proceedings \citealt{Ibgui_et_al_2012_3,Ibgui_et_al_2012_4}; we are also writing a paper on this topic: Ibgui et al. 2012,
in preparation).

We envision various extensions in the near future. First, we will implement an iterative method based on the
Accelerated Lambda Iteration (ALI) technique to handle scattering and NLTE effects.
We will interface IRIS with the NLTE stellar atmosphere code TLUSTY \citep{Hubeny_1988,Hubeny_and_Lanz_1995},
from which we take the treatment of atomic data, and all local physics (opacities, and the scheme for
solving the set of statistical equilibrium equations using the preconditioning method to obtain NLTE
level populations). Later, extending IRIS to polarized radiative transfer will provide the code with
the capability to diagnose magnetic fields.

On the computational side, IRIS is written in Fortran 95. The current version runs on a single processor. We plan to
parallelize the code with the message passing interface (MPI) library. 
Post-processing of (R)(M)HD calculations performed on adaptively refined grids generated with the adaptive mesh refinement (AMR) technique
will benefit from 3D nonuniform Cartesian grids in IRIS, though its full implementation will require additional work.

\begin{acknowledgements}
We are grateful to Matthias Gonz{\'a}lez (AIM, CEA/DSM/IRFU, CNRS, Universit{\'e} Paris Diderot, 91191 Gif-sur-Yvette, France) for providing
us with the 3D hydrodynamics results generated by the HERACLES code. These simulations, which mimic a 1D radiative shock in ideal gas, were used
as an input for the radiative transfer code IRIS, in our test to reproduce the radiative field in a 1D plane-parallel medium (see 
Section~\ref{subsec:cmp_3D_1D}).
We are also grateful to the referee for a prompt report and insightful suggestions that have improved the paper.
The work is supported by French ANR, under grant 08-BLAN-0263-07.
The work of IH was supported in part by NSF grant AST-0807496.
IH also acknowledges the travel support from the Observatoire de Paris, and the Visiting Professorship at the
Universit{\'e} Pierre et Marie Curie where a part of work was done.
\end{acknowledgements}

\bibliographystyle{aa}
\bibliography{biblio}

\end{document}